\documentclass[aps,prl,twocolumn,epsfig,showpacs]{revtex4}
\usepackage[centertags]{amsmath}
\usepackage{amssymb}
\usepackage{amsthm}
\usepackage{graphicx,subfigure}
\usepackage{epsfig}

\begin{document}

\title{Super-Critical and Sub-Critical Hopf bifurcations in two and three dimensions}

\author{Debapriya Das$^1$, Dhruba Banerjee$^1$ and Jayanta K. Bhattacharjee$^2$}

\affiliation{
$^1$ Department of Physics, Jadavpur University, Kolkata, India \\
$^2$ Harish Chandra Research Institute, Allahabad, India.}

\maketitle

\section*{ABSTRACT}
Hopf bifurcations have been studied perturbatively under two broad headings, viz., super-critical and sub-critical. The criteria for occurrences of such bifurcations have been investigated using the renormalization group. The procedure has been described in details for both two and three dimensions and has been applied to several important models, including those by Lorenz and Rossler.\\
\section*{I:INTRODUCTION}

Hopf bifurcations, introduced quantitatively in the next section, have played a pivotal role in the development of the theory of dynamical systems in different dimensions \cite{nayfeh,gia,san,bog,min,jord,rand,stro,laksh,forest}. The uniqueness of such bifurcations lies in two aspects: unlike other common types of bifurcations (viz., pitchfork, saddle-node or transcritical) Hopf bifurcations cannot occur in one dimension. The minimum dimensionality has to be two. The other aspect is that Hopf bifurcations deal with birth or death of a limit cycle (LC) as and when the LC emanates from or shrinks onto a fixed point, the focus. Thus, unlike the other kinds of bifurcations which mostly deal with stability properties of (fixed) points, Hopf bifurcations deal with points as well as isolated phase orbits, the limit cycles. Two types of Hopf bifurcations are common and they go under the broad headings: super-critical (forward) and sub-critical (backward)\cite{bog,min,jord,rand,stro,guo,basso}(defined in Sec.\textbf{II}). In this paper the criteria for occurrences of these two types of Hopf bifurcations have been studied using renormalization group \cite{deb,amartya,cgo,paq,db,dhd,guck}, the operational aspects of which have been elaborately explained both for two and three dimensions. Although there exists a criterion \cite{stro,guck} that deals with such aspects in two dimensional dynamical systems, in three dimensions, no definite method for deciding super-critical or sub-critical Hopf bifurcations exists. We propose that the renormalization group (RG) procedure can be used to adress the issue of forward or backward in higher dimensions if at the instability point the eigenvalues are all negative except for two which are a pair of imaginary numbers.

\noindent The paper has been organized as follows: in Sec.\textbf{II} we rederive the well-known criterion that is commonly used to discriminate super-critical and sub-critical Hopf bifurcations, using the RG. We also explain through an example how the predictions made by the RG have certain advantages. In Sec.III we develop the RG-procedure for three dimensions. Sec.\textbf{IV} and \textbf{V} are devoted to detailed analyses of Hopf bifurcations in the Lorenz and Rossler models \cite{guck,li,liu,ding}respectively, where the formalism developed in Sec.\textbf{III } has been extensively applied. The paper has been summarized in Sec.\textbf{VI}.

\section*{II:RG IN 2D HOPF-BIFURCATIONS}
In this section we first introduce Hopf-bifurcations briefly for two-dimensional dynamical systems followed by a detailed analysis of how the amplitude equation (derived from the RG) can be used to understand its super-critical or sub-critical nature.\\
\noindent A 2D-dynamical system, which in polar form looks like\\

\begin{equation}
\dot{r}=\mu r - \lambda r^{3}\label{eq1}
\end{equation}
\begin{equation}
\dot{\theta}=\omega\label{eq2}
\end{equation}
\noindent undergoes Hopf-bifurcation when the co-efficient of the linear term of Eq.(\ref{eq1}), i.e. $\mu$, becomes zero. The bifurcation is super critical if $\lambda>0$ and subcritical if $\lambda<0$ When $\mu$ $>0$ i.e. the origin is an unstable spiral.
\noindent For an arbitrary two dimensional system it is non-trivial to establish whether a Hopf bifurcation is forward or backward. There is a well established criterion [Guckenheimer/Holmes] that decides which way the system will go and the method used to arrive at it uses centre manifold theory. Here we shall see how the Renormalization Group comes to our help in deciding whether for a generalized 2D system undergoing Hopf bifurcation,it will be super critical or sub critical. The result that we arrive at perturbatively using the RG is the same as that obtained by centre-manifold theory. Hence, this section serves as a good rehearsing ground for applying the RG-technique, which has been employed to study Hopf bifurcations in 3D in the next section. Therefore let us start out with a time-scaled ($\tau=\omega t$) 2D-dynamical system \cite{stro},
\begin{equation}
\dot{x}=\mu x - y +\lambda f(x,y)\label{eq3}
\end{equation}
\begin{equation}
\dot{y}= x+\mu y +\lambda g(x,y)\label{eq4}
\end{equation}
\noindent where to effect a perturbation analysis, we have taken the nonlinear parts included in the functions $f(x,y)$ and $g(x,y)$ as small, $\lambda$ being the perturbation parameter. The polynomial structures of these nonlinear functions can be written as\\
\begin{equation}
f(x,y)=\sum_{i,j}f_{ij} x^{i} y^{j}~~~~~ (i+j\geq 2)\label{eq5}
\end{equation}
\noindent and
\begin{equation}
g(x,y)=\sum_{i,j} g_{ij} x^{i} y^{j}~~~~~~(i+j\geq 2).\label{eq6}
\end{equation}
\noindent Differentiating Eq.(\ref{eq3}) with respect to time, we get\\
\begin{equation}
\ddot{x}=\mu \dot{x} - \dot{y} + \lambda \left[\dot{x} \frac{df}{dx} + \dot{y} \frac{df}{dy}\right]\nonumber
\end{equation}
\begin{eqnarray}
\Rightarrow \ddot {x} + (1 - \mu^{2})x &=& -2\mu y\nonumber\\
 &+& \lambda \left[ \mu f(x,y) - g(x,y)+ \mu x \frac{df}{dx}\right.\nonumber\\
&-& \left.y \frac{df}{dx} + x \frac{df}{dy} - \mu y \frac{df}{dy}\right]\nonumber\\
&+& \lambda^{2} \left[f(x,y) \frac{df}{dx} + g(x,y) \frac{df}{dy}\right].\label{eq7}
\end{eqnarray}
\noindent We see that Hopf bifurcation occurs right at the point $\mu=0$ and that the origin is unstable for $\mu>0$. To analyze the role of the lowest nonlinear term in driving the system at that point, we put $\mu=0$ in the above equation to obtain\\
\begin{equation}
\ddot{x} + x=\lambda \left[-g(x,y) -y \frac{df}{dx} + x\frac{df}{dy}\right].\label{eq8}
\end{equation}
\noindent Here $\lambda$ being a perturbation parameter we can expand $x$ and $y$ perturbatively as\\
\begin{equation}
x=x_{0} + \lambda x_{1} + \lambda^{2} x_{2}+......\label{eq9}
\end{equation}
\begin{equation}
y=y_{0} + \lambda y_{1} + \lambda^{2} y_{2}+....\label{eq10}
\end{equation}
\noindent The RG-technique, which we apply here to derive the amplitude and phase equations has been discussed in details in \cite{db,amartya,deb}. The central idea lies in `cutting-off' the secular divergences arising from integration of the resonant terms, by introducing a flexible origin of the time scale. This flexibility in the choice of the origin leads to the RG-flow equations, which appear in the guise of the amplitude and phase equations of the problem. The result is that at the $n^{th}$ order of perturbation, the equation\\
\begin{eqnarray}
\ddot{x_{n}}+ \omega^{2} x_{n}&=& P_{n}(a)\sin(\omega t+\theta) + Q_{n}(a)\cos(\omega t+\theta)+\nonumber\\ &&\mathrm{~other~regular~(non-resonant)~terms~of~}\nonumber\\
&&\mathrm{lower~orders~in~perturbation}\label{eq11}
\end{eqnarray}
\noindent where $P_{n}(a)$ and $Q_{n}(a)$ are functions of the amplitude `$a$', leads to the amplitude and phase equations as,\\
\begin{equation}
\frac{da}{dt}=-\frac{\lambda^{n}P_{n}}{2\omega} + \mathrm{lower~order~terms~in~\lambda}\label{eq12}
\end{equation}
\begin{equation}
\frac{d\theta}{dt}=-\frac{\lambda^{n}Q_{n}}{2a\omega}+\mathrm{lower~order~terms~in~\lambda}.\label{eq13}
\end{equation}
\noindent This result may seem similar to that derived by standard perturbative techniques like averaging or multiple-time-scale analysis, there are subtle differences \cite{cgo} between these methods and the RG, which, however, will not concern us in the discussions to follow. Our objective here will be to write the amplitude equation for Eq.(\ref{eq8}) upto a relevant order of perturbation so that we can understand the role of the $a^{3}$-term (lowest nonlinear power of `$a$'), in governing the dynamics. By `relevant order' we mean, that, beyond that order of perturbation there cannot be any `$a^{3}$'-term, in the amplitude equation. Therefore, in what follows, our quest will be to identify the $a^{3}\sin(t+\theta)$ terms from the RHS of Eq.(\ref{eq8}). With the lowest power of $x$ and $y$ in $f(x,y)$ and $g(x,y)$ as 2, it is to understand that third and higher orders of perturbation will not contain $a^{3}$-terms. That is why perturbative calculations upto second order suffice our purpose. Here state the main expressions only and have shown all the steps in the appendix.\\

\noindent Returning to Eq.(\ref{eq8}), we can Taylor-expand the functions on the RHS by involving the perturbation expansions Eq.(\ref{eq13}) and Eq.(\ref{eq14}) upto order $O(\lambda^{2})$ to get the following equations$:-$\\

\begin{equation}
\ddot{x_{0}} + x_{0}=0\label{eq14}
\end{equation}
\begin{equation}
\ddot{x_{1}} + x_{1} = -g(x_{0},y_{0}) - y_{0}\frac{df}{dx}(x_{0},y_{0}) + x_{0} \frac{df}{dy}(x_{0},y_{0})\label{eq15}
\end{equation}
\begin{eqnarray}
\ddot{x_{2}}+ x_{2}=&-&x_{1} \frac{\partial g}{\partial x}(x_{0},y_{0}) + x_{1}\frac{\partial f}{\partial y}(x_{0},y_{0})\nonumber\\
&-&y_{1}\frac{\partial g}{\partial y}(x_{0},y_{0})-y_{1} \frac{\partial f}{\partial x}(x_{0},y_{0})\nonumber\\
 &-& y_{0}x_{1}\frac{\partial^{2}f}{\partial x^{2}}(x_{0},y_{0})-y_{0}y_{1} \frac{\partial^{2}f}{\partial x\partial y}(x_{0},y_{0})\nonumber\\
 &+&x_{0}x_{1}\frac{\partial^{2}f}{\partial x\partial y}(x_{0},y_{0}) + x_{0}y_{1}\frac{\partial^{2}f}{\partial y^{2}}(x_{0},y_{0})\nonumber\\
&+& f(x_{0},y_{0})\frac{\partial f}{\partial x}(x_{0},y_{0}) + g(x_{0},y_{0})\frac{\partial f}{\partial y}(x_{0},y_{0}).\nonumber\\\label{eq16}
\end{eqnarray}
\noindent Writing the solution of Eq.(\ref{eq14}) as\\
\begin{equation}
x_{0}=a\cos( t+\theta)=a\cos \phi \mathrm{~(for~short)}\label{eq17}
\end{equation}
\noindent we have from Eq.(\ref{eq3}) (with $\mu=0$)\\
\begin{equation}
y_{0}=-\dot{x}_{0}=a\sin(t +\theta)=a\sin \phi.\label{eq18}
\end{equation}
\noindent First-order calculations lead to\\
\begin{equation}
\ddot{x_{1}}+x_{1}=\mathrm{~regular~terms}-\frac{a^{3}}{8}(g_{xxy}+g_{yyy} + f_{xxx} + f_{xyy})\sin\phi.\label{eq19}
\end{equation}
\noindent In the second order calculations (Eq.(\ref{eq16})) we have to know the renormalized expressions of $x_{1}$ and $y_{1}$.
For this, it is important to realize from Eq.(\ref{eq15}) and from\\
\begin{equation}
y_{1}= -\dot{x}_{1} + f(x_{0},y_{0})\label{eq20}
\end{equation}
\noindent that the lowest power of `$a$' in $x_{1}$ and $y_{1}$ is 2. Therefore the regular (non-secular) $a^{2}$-terms on the RHS of Eq.(\ref{eq15}) have to be identified in order to get the $a^{3}$-terms of Eq.(\ref{eq16}) that contribute to the amplitude equation. Considerations of this kind lead to the following (details are done in the Appendix)$:-$\\

\begin{eqnarray}
x_{1}&=&-a^{2} \left[\frac{1}{4}(g_{xx} + g_{yy}) + \frac{1}{12}(4f_{xy} - g_{xx} + g_{yy}) \cos2\phi \right.\nonumber\\
&+& \left.\frac{1}{6}(-f_{xx} + f_{yy} - g_{xy})\sin2\phi\right] \nonumber\\
&+& \mathrm{higher~powers~of ~}a.\label{eq21}
\end{eqnarray}
\noindent Accordingly, from Eq.(\ref{eq20}), we get the renormalized $y$ as\\
\begin{eqnarray}
y_{1}&=&a^{2}\left[\frac{1}{4}(f_{xx} + f_{yy}) +\frac{1}{12}(-f_{xx} +f_{yy} -4g_{xy})\cos2\phi\right.\nonumber\\
&+&\left.\frac{1}{6}(-f_{xy} +g_{xx} -g_{yy})\sin2\phi\right]\nonumber\\
&+& \mathrm{higher~powers~of~}a.\label{eq22}
\end{eqnarray}

\noindent This leads to\\

\begin{eqnarray}
\ddot{x_{2}} + x_{2} &=& \frac{a^{3}}{8}\sin\phi \left[g_{xy}(g_{xx} + g_{yy}) - f_{xy}(f_{xx}+f_{yy})\right.\nonumber\\
&+&\left. (f_{xx}g_{xx}-f_{yy}g_{yy})\right]+\mathrm{ regular~terms}.\label{eq23}
\end{eqnarray}
\noindent Finally, using the general result of Eq.(\ref{eq12}), we get the amplitude equation upto second order as [combining Eqs.(\ref{eq19}) and Eq.(\ref{eq23})],\\
\begin{eqnarray}
\frac{da}{dt}&=&\frac{a^{3}}{16}\left[\lambda(f_{xxx}+ f_{xyy} +g_{xxy} + g_{yyy})\right.\nonumber\\
&+&\left.\lambda^{2} (f_{xy}(f_{xx} + f_{xy})-g_{xy}(g_{xx}+g_{yy})\right.\nonumber\\
&-&\left. f_{xx}g_{xx} + f_{yy}g_{yy})\right]+ \mathrm{higher~powers~of~}a.\nonumber\\ \label{eq24}
\end{eqnarray}
\noindent The sign of the quantity within the $[~]$ brackets dictate the dynamics right at the point of the Hopf bifurcation ($\mu=0$ in Eq.(\ref{eq3}) and Eq.(\ref{eq4})). If the sign of this quantity be negative, then the nonlinear amplitude term  of lowest power (here $a^{3}$) drives the system towards the origin and we get a super-critical Hopf bifurcation. On the contrary, when the sign is positive, this nonlinear term drives the system away from the origin which is the case of sub-critical Hopf bifurcation.
\noindent As an example, let us consider the vanderpol equation,\\
\begin{equation}
\dot{x}=-y\label{eq25}
\end{equation}
\begin{equation}
\dot{y}=x-(x^{2}-1)y.\label{eq26}
\end{equation}

\noindent This is a Lienard system where the function $f(x,y)$ [see Eq.(\ref{eq3})] is zero (and also $\mu=0$). The criterion of Eq.(\ref{eq24}) (with $\lambda=1$)evaluates to\\
\begin{equation}
\frac{da}{dt}=\frac{a^{3}}{16}(-1)\label{eq27}
\end{equation}
\noindent with all the partial derivatives evaluated at $x=y=0$. This implies supercritical Hopf bifurcation at $\mu=0$ which means, the lowest-power nonlinear term drives the system towards the origin.\\
\noindent Now, let us apply the same principle to the oscillator\\
\begin{equation}
\dot{x}=-y\label{eq28}
\end{equation}
\begin{equation}
\dot{y}=x-(x^{2}-\alpha)(x^{2}-\beta)y^{3}.\label{eq29}
\end{equation}
\noindent In this case, the criterion of Eq.(\ref{eq24}) tells us that\\
\begin{equation}
\frac{da}{dt}=-\frac{\alpha\beta}{16}\label{eq30}
\end{equation}
\noindent which means that for ($\alpha, \beta$) of the same sign, the bifurcation will be supercritical and for ($\alpha, \beta$) of opposite signs it will be subcritical. But, this is just the reverse to what one sees in the numerical phase plot. Numerically, it is seen that when ($\alpha, \beta$) are both positive then there is a stable origin girdled by an unstable  limit cycle (LC) which in turn is surrounded by a stable LC. As $\beta$ is made zero, the inner LC gradually engulfs the stable origin rendering it unstable, which is clearly a case of subcritical Hopf bifurcation, as opposed to the super-critical case predicted by Eq.(\ref{eq24}). This is because the fixed point $x=y=0$ is not an unstable spiral but a centre for all $\alpha$ and $\beta$ and hence Eq.(\ref{eq24}) is not applicable. Parturbative RG can still be employed by combining Eq.(\ref{eq28})and Eq.(\ref{eq29}) as\\
\begin{equation}
\ddot{x}+x=-\lambda(x^{2}-\alpha)(x^{2}-\beta)\dot{x}^{3}\label{eq31}
\end{equation}

\noindent (where $\lambda$ is a perturbation parameter) and then, just by identifying the coefficients of $\sin(t+\theta)$ from the RHS of Eq.(\ref{eq31}), we arrive at the amplitude equation\\
\begin{equation}
\frac{da}{dt}=-\frac{a^{3}}{128}[48\alpha\beta-8(\alpha+\beta)a^{2}+3a^{4}]\label{eq32}
\end{equation}

\noindent which has the fixed points at
\begin{eqnarray}
&&a=0\mathrm{~~and~~}\nonumber\\
&&a=[\frac{4}{3}\{(\alpha+\beta)\pm\sqrt{(\alpha+\beta)^{2}-9\alpha\beta}\}]^{\frac{1}{2}}.\label{eq33}
\end{eqnarray}

\begin{figure}
\subfigure[]{\includegraphics[width=0.50\textwidth,angle=0,clip]{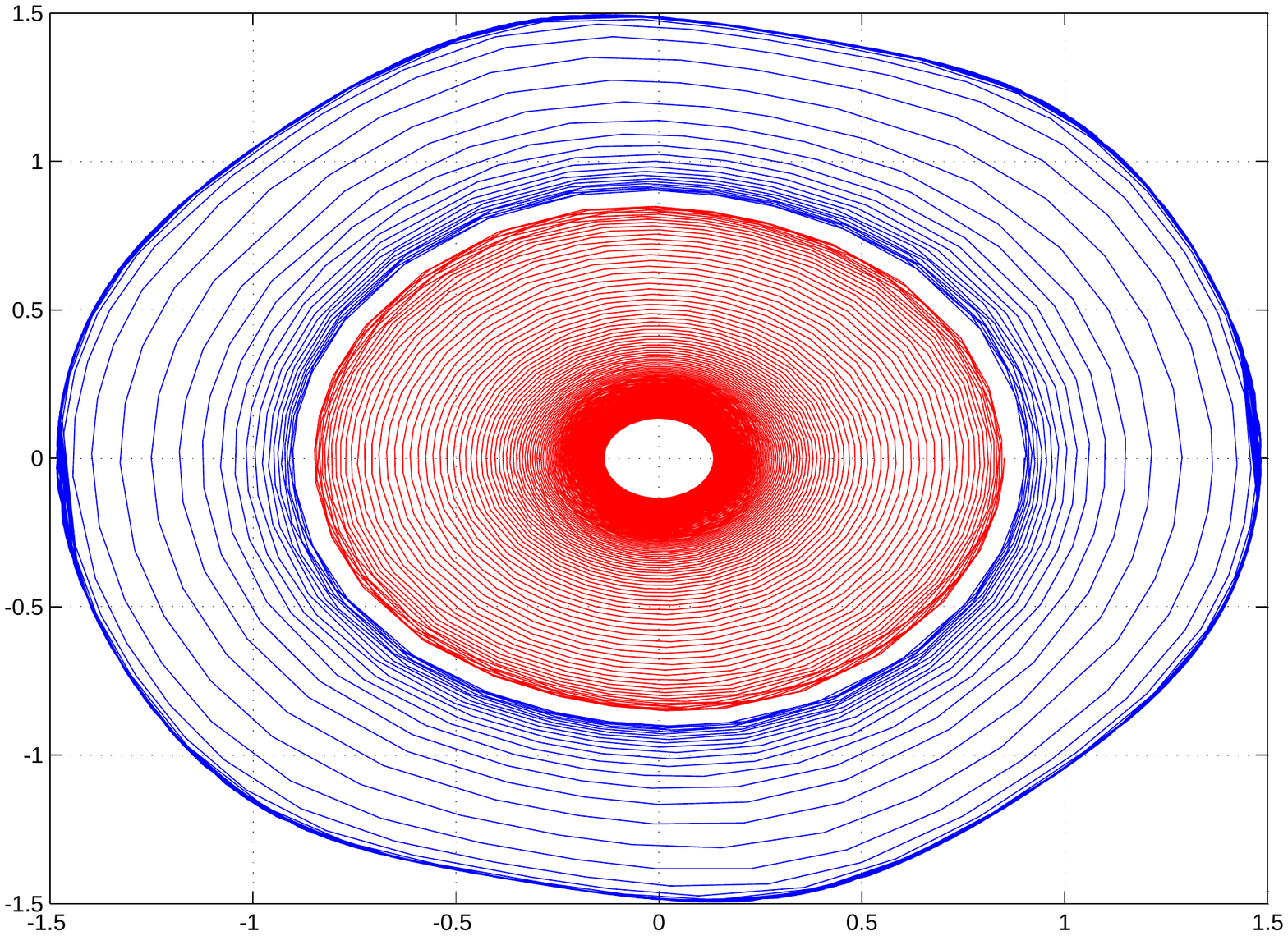}}\\
\subfigure[]{\includegraphics[width=0.50\textwidth,angle=0,clip]{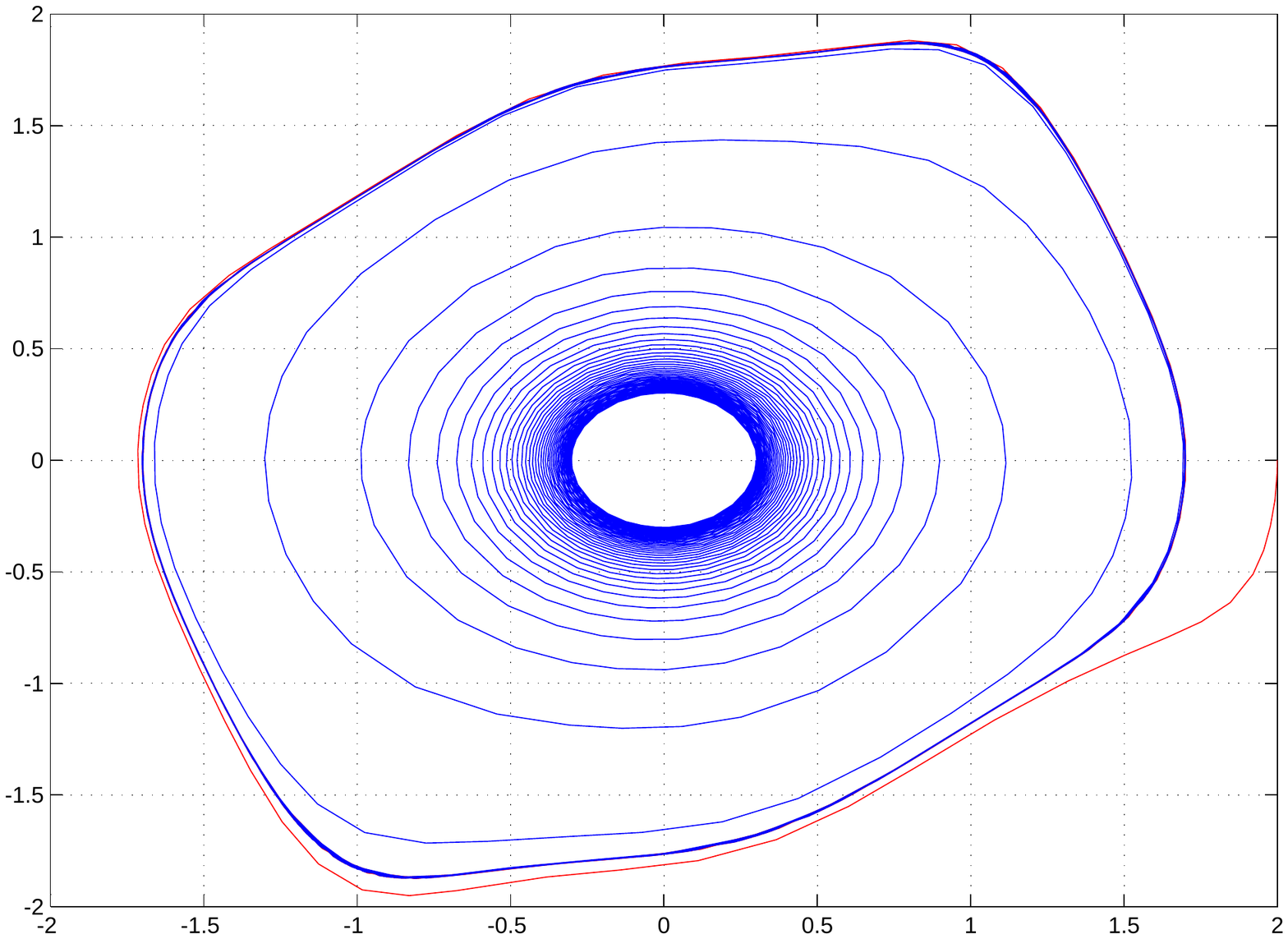}}\\
\caption[]{Phase plots of the system $\ddot{x} + x + {\dot{x}}^3 (x^2 -\alpha) (x^2 - \beta)$ are shown, with $\beta$ held at $1$, a): $\alpha=0.1$ makes the origin stable surrounded by an inner unstable LC(at 0.86) and outer stable LC(at 1.47). The inner LC shrinks to $0$ at $\alpha=0$ in a subcritical Hopf bifurcation. (b) Here $\alpha = -0.1$ gives an unstable origin girdled by a stable LC at $a=1.7$. The locations of the respective LCs are exactly predicted by the fixed points of the amplitude equation (see Eq.(\ref{eq65})).}
\end{figure}

\begin{figure}
\subfigure[]{\includegraphics[width=0.50\textwidth,angle=0,clip]{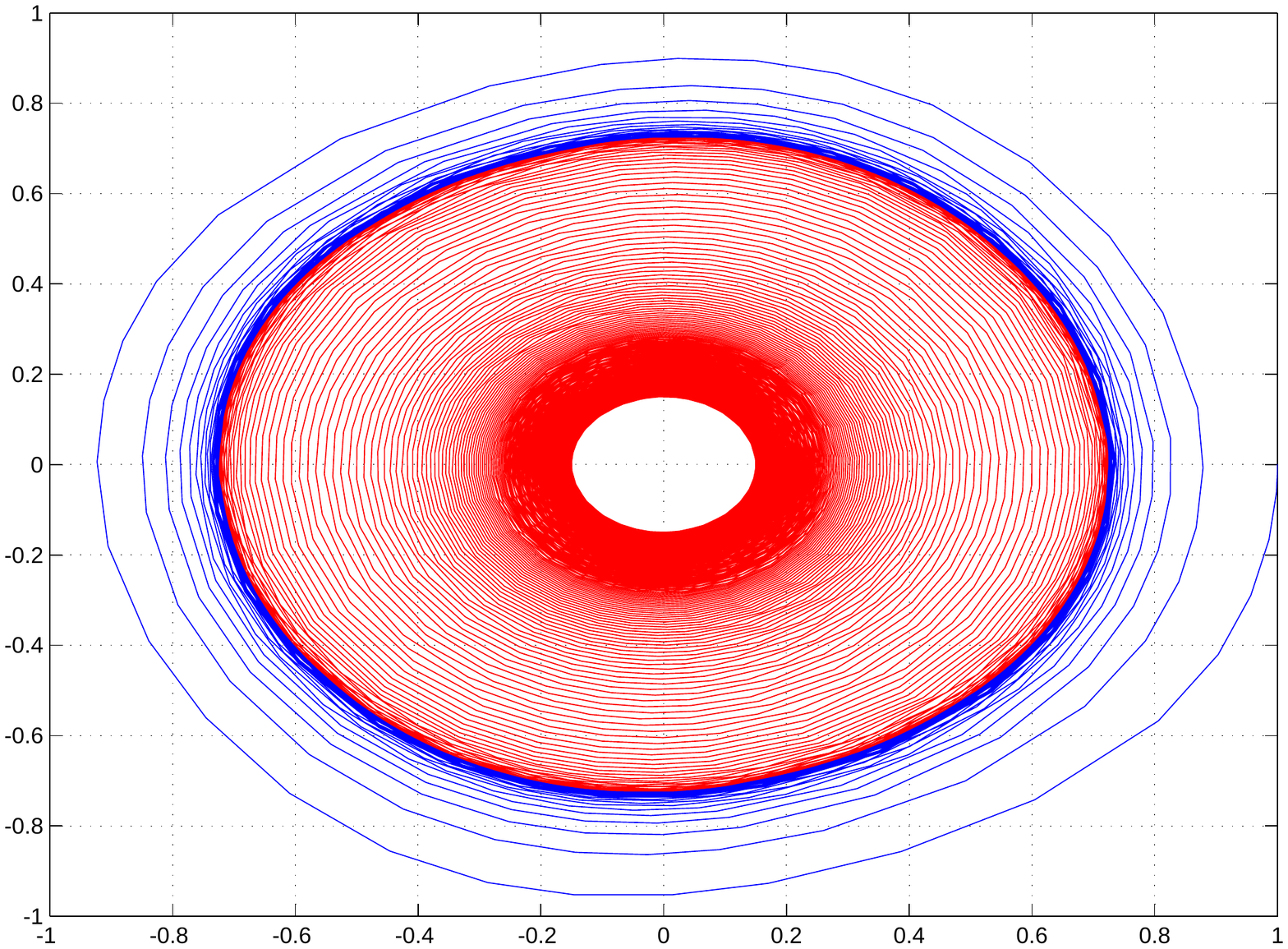}}\\
\subfigure[]{\includegraphics[width=0.50\textwidth,angle=0,clip]{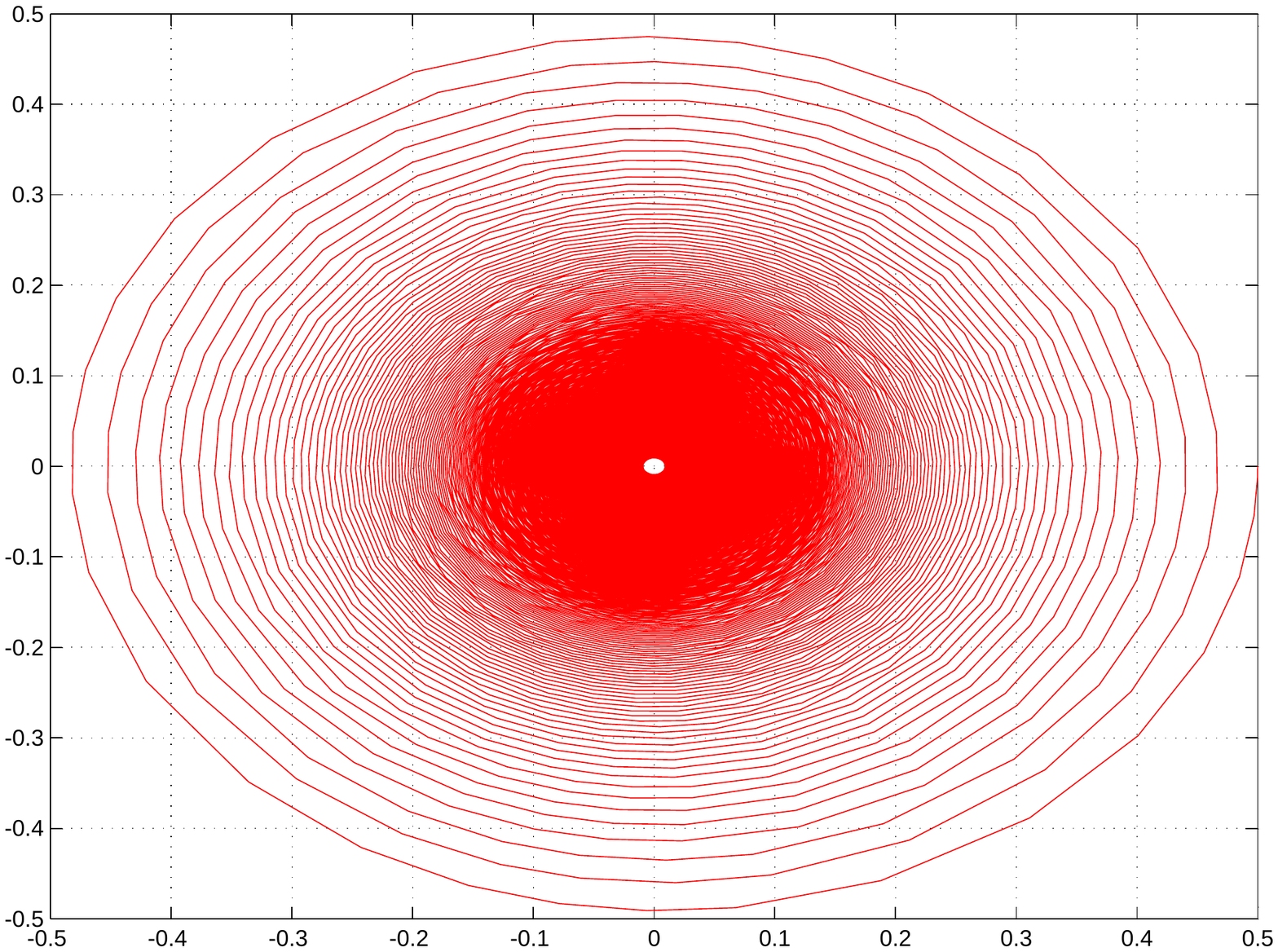}}
\caption[]{Phase plots of the system $\ddot{x} + x + {\dot{x}}^3 (x^2 -\alpha) (x^2 - \beta)$ are given. Keeping $\beta=-1$, (a) $\alpha=-0.1$ makes the origin unstable and gives one stable LC at $a=0.72$, which shrinks to $0$ at $\alpha=0$ in a supercritical Hopf bifurcation. (b) Here $\alpha = 0.1$ giving a stable origin with no LC.around. The locations of the respective LCs are found exactly as predicted by the fixed points of the amplitude equation (see Eq.(\ref{eq65})).}
\end{figure}

\noindent From Eq.(\ref{eq32}) and Eq.(\ref{eq33}), it is easy to see that for ($\alpha,\beta>0$), the origin is stable surrounded by two LCs, inner unstable and outer stable. As $\beta\rightarrow0$, the system undergoes a subcritical Hopf bifurcation corroborated accurately by the plot of Fig.1. With $\alpha>0$ and $\beta<0$, Eq.(\ref{eq32}) and Eq.(\ref{eq33}) tell us that there is an unstable origin girdled by a stable LC. This is the scenario of a Super-critical Hopf bifurcation that occurs as $\alpha\rightarrow0$ [see Fig.2].\\

\section*{III:RENORMALIZATION GROUP IN THREE DIMENSIONS}

In this Section we explicitly show how the RG-procedure works in 3D. Before going into the specific systems (Lorenz and Rossler models) let us consider a differential equation of the form\\
\begin{equation}
f(D)u= R \cos \omega t + S \sin \omega t\label{eq34}
\end{equation}

\noindent where $f(D)$ is some cubic polynomial of the differential operator $D\equiv \frac{d}{dt}$ and is factorizable as\\
\begin{equation}
f(D)=(D^{2}+\omega^{2})(D+\alpha).\label{eq35}
\end{equation}
\noindent Here $\alpha$ is some number and $\omega$ is the same frequency that occurs in the resonant terms on the RHS of  Eq.(\ref{eq34}). As we shall see in the next two sections, that a differential equation of the form of  Eq.(\ref{eq34}) emerges naturally in the study of dynamical systems like the Lorenz or Rossler attractors. On integrating  Eq.(\ref{eq34}) we get at the first stage,\\
\begin{eqnarray}
(D^{2}+\omega^{2})u &=& \left[\frac{R}{-\omega^{2}-\alpha^{2}} \cos \omega t \right.\nonumber\\
&& \left.+ \frac{S}{-\omega^{2}-\alpha^{2}} \sin \omega t \right]\nonumber\\
&=&\left[ P \cos \omega t + Q \sin \omega t\right]\label{eq36}
\end{eqnarray}
\noindent where in $P$ and $Q$, the co-efficients of $\cos \omega t$ and $\sin \omega t$ from the RHS of  Eq.(\ref{eq34}) get mixed up (unlike the 2D case) as,\\
\begin{equation}
P=(R \alpha +S \omega)/( \omega^{2} + \alpha^{2})\label{eq37}
\end{equation}
\begin{equation}
Q=(R \omega + S \alpha)/(\omega^{2} + \alpha^{2})\label{eq38}
\end{equation}

\noindent Another integration yields\\
\begin{eqnarray}
u &=& \frac{1}{4\omega^{2}}\left[P \cos \omega t + Q \sin \omega t \right]\nonumber\\
 &&+ \frac{t}{2\omega} \left[P \sin \omega t - Q \cos \omega t\right].\label{eq39}
\end{eqnarray}
\noindent The $t$-divergences in the last two terms of  Eq.(\ref{eq39}) are physically unacceptable in a perturbation theory where `$u$' plays the role of some `$x_{n}$' in a perturbative expansion of the form of  Eq.(\ref{eq9}), (viz., $x=x_{0}+\lambda x_{1} + \lambda^{2} x_{2}+....$) built around some purely oscillatory form of $x_{0}$ given by $x_{0}= A \cos \omega t + B \sin \omega t$. Then the RG-method consists of expanding $A$ and $B$ perturbatively and defining renormalization constants $Z_{n}^{(A)}$ and $Z_{n}^{B}$ so as to cut-off the secular divergences order by order. Thus,\\
\begin{eqnarray}
A&=&A(\mu)\left[1+ \lambda Z_{1}^{(a)} + \lambda^{2}Z_{2}^{(a)}+.....\right]\label{eq40}\\
B&=&B(\mu)\left[1+\lambda Z_{1}^{(b)} + \lambda^{2}Z_{2}^{(b)}+......\right]\label{eq41}
\end{eqnarray}
\noindent where `$\mu$' is a new arbitrary time origin introduced to sieve out a regularized part from the $t$-divergent terms of  Eq.(\ref{eq39}) as, $t \sin \omega t=(t-\mu+\mu)\sin \omega t$, and a similar split-up for the $t \cos \omega t$ term. If we are working at the $n^{th}$ order of perturbation, then this split up along with Eq.(\ref{eq40}) and  Eq.(\ref{eq41}) allow us to define the renomalization constants in such a way as to nullify the divergent $\mu \sin \omega t$ and $\mu \cos \omega t$ terms order by order. Thus, at the $n^{th}$ order of perturbation we define\\
\begin{equation}
Z_{n}^{(a)}=\frac{\mu}{A(\mu)}\frac{Q}{2\omega}\label{eq42}
\end{equation}

\noindent and\\
\begin{equation}
Z_{n}^{(b)}=-\frac{\mu}{B(\mu)}\frac{P}{2\omega}\label{eq43}
\end{equation}

\noindent where, it is presumed that the divergences upto $(n-1)^{th}$ order of perturbation have already been similarly renormalized.\\
\noindent This leaves us with the result\\
\begin{eqnarray}
u&=&A(\mu) \cos \omega t + B(\mu) \sin \omega t \nonumber\\
&&+ \mathrm{renormalized~lower~order~terms}\nonumber\\
&&+\lambda^{n} \left[(t-\mu) \frac{P}{2\omega} \sin \omega t - (t-\mu) \frac{Q}{2\omega} \cos \omega t\right.\nonumber\\
&& \left.+ \mathrm{other~non-divergent~terms} \right].\label{eq44}
\end{eqnarray}

\noindent Now, since the time origin $\mu$ was chosen as arbitrary, therefore the dynamics should be independent of $\mu$. Thus,\\
\begin{eqnarray}
\frac{da}{d\mu}=0&=& \frac{dA}{d\mu} \cos \omega t+ \frac{dB}{d\mu} \sin \omega t\nonumber\\
&&+ \lambda^{n} \left[-\frac{P}{2\omega} \sin \omega t+ \frac{Q}{2\omega} \cos \omega t\right]\label{eq45}
\end{eqnarray}
\noindent whence, equating the co-efficients of $\cos \omega t$ and $\sin \omega t$ terms give the RG-flow equations at the $n^{th}$ order as\\
\begin{equation}
\frac{dA}{d\mu}=-\lambda^{n} \frac{Q}{2\omega} + \mathrm{lower~order~terms~in~}\lambda\label{eq46}
\end{equation}

\begin{equation}
\frac{dB}{d\mu}=\lambda^{n} \frac{P}{2\omega}+ \mathrm{lower~order~terms~in~}\lambda.\label{eq47}
\end{equation}

\noindent Finally, exploiting the arbitrariness of $\mu$, we put $\mu=t$ in Eq.(\ref{eq44}) to obtain the renormalized $u(t)$ upto the $n^{th}$ order. Thus, the central results transpiring from the divergent terms of Eq.(\ref{eq39}) are\\
\begin{equation}
\frac{dA}{dt}=\lambda^{n} \frac{1}{2\omega}\left[\mathrm{coefficient~of~}t\cos \omega t\right]+\mathrm{lower~ order~ terms ~in}\lambda\label{eq48}
\end{equation}
\begin{eqnarray}
\frac{dB}{dt}&=&\lambda^{n}\frac{1}{2\omega}\left[\mathrm{coefficient~of~}t\sin \omega t\right]\nonumber\\&+&\mathrm{lower~order~terms~in} \lambda.\label{eq49}
\end{eqnarray}

\noindent We shall require these results in the following two Sections devoted to the study of super-critical and sub-critical Hopf bifurcations in the Lorenz and Rossler systems.

\section*{IV:LORENZ MODEL}
In this Section we shall use the RG-method described in \textbf{Sec.III} to study super-critical and sub-critical Hopf bifurcations in one of the most important and historically famous 3D-dynamical systems, viz. the Lorenz attractor, described by the equations,
\begin{eqnarray}
\dot{x}&=&\sigma (-x+y)\label{eq50}\\
\dot{y}&=& -xz + rx -y\label{eq51}\\
\dot{z}&=& xy -bz\label{eq52}
\end{eqnarray}

\noindent where $b$, $r$ and $\sigma$ are controllable parameters, all positive. This system, has its fixed point at $x_{0}=y_{0}=\sqrt{b(r-1)}$ and $z_{0}=r-1$. Therefore, shifting the origin to the fixed point we move to a new set of co-ordinates $X=x-\alpha$, $Y=y-\alpha$ and $Z=z-(r-1)$ where\
\begin{equation}
\alpha=\sqrt{b(r-1)}.\label{eq53}
\end{equation}

\noindent This leads to a new set of equations in the shifted co-ordinates as
\begin{eqnarray}
(D+\sigma)X &=& \sigma Y \label{eq54}\\
(D+1)Y&=&-XZ+X- \alpha Z\label{eq55}\\
(D+b)Z&=&XY+ \alpha X + \alpha Y\label{eq56}
\end{eqnarray}

\noindent where, as before, $D\equiv\frac{d}{dt}$. From this point, our focus will be to cast this system in the form of Eq.(\ref{eq34}) and Eq.(\ref{eq35}) of the last Section. In doing so, we note that Eq.(\ref{eq55}) can be written as,\\
\begin{equation}
Z=- \frac{1}{\alpha}\{(D+1)Y +XZ-X\}\label{eq57}
\end{equation}

\noindent which when placed in Eq.(\ref{eq56}) yields \\
\begin{equation}
(D+b){(D+1)Y+XZ-X}=-\alpha XY- \alpha^{2} X-\alpha^{2}Y.\label{eq58}
\end{equation}

\noindent Substituting $Y$ form Eq.(\ref{eq54}) in Eq.(\ref{eq58}) we have, after some algebra and rearrangements,\\
\begin{eqnarray}
D^{3}X&+&(1+\sigma+b)D^{2}X+ \omega_{0}^{2} DX + \omega_{0}^{2}(1+\sigma+b)X\nonumber\\
&=& -\sigma (D+b)XZ-\alpha\sigma XY +(r_{0}-r)b\dot{X}\nonumber\\
&+&2 \sigma b(r_{0}-r)X\label{eq59}
\end{eqnarray}
\noindent where\\
\begin{eqnarray}
r_{0}&=&\sigma \frac{\sigma+b+3}{\sigma-b-1}\label{eq60}\\
\omega_{0}^{2}&=&\frac{2b\sigma(1+\sigma)}{\sigma-b-1}.\label{eq61}
\end{eqnarray}

\noindent We note that the LHS of Eq.(\ref{eq59}) is already in the form of that of Eq.(\ref{eq34}). The operator on the LHS of Eq.(\ref{eq59}) is factorizable as\\
\begin{equation}
LX= (D^{2} + \omega_{0}^{2})(D+\sigma+b+1)X. \label{eq62}
\end{equation}

\noindent From this point to make progress, we take a perturbative approach by tagging the RHS of Eq.(\ref{eq59}) with some perturbation parameter $\varepsilon$ and invoking the perturbative expansions in $X$, $Y$ and $Z$ as \\
\begin{equation}
X=X_{0}+\varepsilon X_{1} +\varepsilon^{2} X_{2}+.........\label{eq63}
\end{equation}

\noindent and similar expansions for $Y$ and $Z$. In the zeroth order the equation\\

\begin{equation}
LX_{0}=0\label{eq64}
\end{equation}

\noindent yields three independent solutions: two trigonometric (oscillatory) and one exponentially decaying. Since we are only interested in long time behaviour we omit the latter from our considerations and with appropriate choice of initial conditions, continue working with the solution\\
\begin{equation}
X_{0}=A \cos \omega_{0}t.\label{eq65}
\end{equation}

\noindent Using this in Eq.(\ref{eq54}) gives the zeroth order for $Y$ as\\

\begin{eqnarray}
Y_{0}&=&X_{0} + \frac{1}{\sigma}\dot{X}_{0}\nonumber\\
&=& A\cos \omega_{0}t- \frac{\omega_{0}}{\sigma} A\sin\omega_{0}t.\label{eq66}
\end{eqnarray}

\noindent Similarly, from the linear terms of Eq.(\ref{eq55}) we get\\
\begin{eqnarray}
Z_{0}&=&\frac{1}{\alpha}(X_{0}-Y_{0}-\dot{Y}_{0})\nonumber\\
&=&\frac{1}{\alpha} \left[(\frac{1}{\sigma}+1) \omega_{0}A\sin\omega_{0} t+ A \frac{\omega_{0}^{2}}{\sigma}\cos \omega_{0} t\right].\nonumber\\
\label{eq67}
\end{eqnarray}

\noindent The $XZ$ term in Eq.(\ref{eq55}) is nonlinear and hence does not participate at this order of calculation. Its explicit presence in the RHS of Eq.(\ref{eq59}) begins to be felt at the first order of perturbation, viz.,\\
\begin{eqnarray}
LX_{1}&=&-\sigma (D+b)X_{0}Z_{0}-\alpha\sigma X_{0}Y_{0} + b(r_{0}-r)\dot{X}_{0}\nonumber\\
&& + 2\sigma b(r_{0}-r)X_{0}. \label{eq68}
\end{eqnarray}
\noindent On using the expressions for $X_{0}$, $Y_{0}$ and $Z_{0}$, Eq.(\ref{eq68}) becomes,\\
\begin{eqnarray}
LX_{1}&=&-\frac{A^{2}}{2\alpha}(\alpha^{2}\sigma + \omega_{0}^{2}b)\nonumber\\
&-& \frac{A^{2}}{2\alpha} \left[2\omega_{0}^{2}(1+\sigma) + \alpha^{2} \sigma + \omega_{0}^{2}b \right] \cos 2\omega_{0}t \nonumber\\
&+& \frac{A^{2}\omega_{0}}{2\alpha}\left[\alpha^{2} - b(1+\sigma) + 2\omega_{0}^{2}\right]\sin 2\omega_{0} t\nonumber\\
&-&2\sigma b \Delta{r} A \cos \omega_{0} t + b\omega_{0} \Delta{r} A\sin\omega_{0}t.\label{eq69}
\end{eqnarray}

\noindent In evaluating $X_{1}$ from the above equation, we note that the first three terms are regular while the last two terms are secular (resonant). On integrating the secular part, we get divergent terms as well as regular $\cos \omega_{0}t$ and $\sin\omega_{0}t$ terms [see Eq.(\ref{eq39})]. Incidentally, these two regular terms do not spawn any further secular terms in the second order calculations which we come to shortly. Hence, for now, we work with the regular part of Eq.(\ref{eq69}) and stack the secular ones to be dealt with along with second-order terms later.
\noindent Introducing constants as\\
\begin{eqnarray}
\beta &=& \sigma+b+1 \label{eq70}\\
P_{1} &=& -\frac{A^{2}}{2\alpha}(\alpha^{2}\sigma + b \omega_{0}^{2}) \label{eq71}\\
P_{2}&=&-\frac{A^{2}}{2\alpha}(\alpha^{2}\sigma + b\omega_{0}^{2} + 2\omega_{0}^{2}(1+\sigma))\label{eq72}\\
P_{3}&=& \frac{A^{2}\omega_{0}}{2\alpha} \left[\alpha^{2} -b(1+\sigma) + 2\omega_{0}^{2}\right].\label{eq73}
\end{eqnarray}

\noindent Eq.(\ref{eq69}) takes the form (with no secular terms in RHS)\\
\begin{equation}
(D^{2} + \omega_{0}^{2})(D+\beta)X_{1} = P_{1} + P_{2} \cos 2\omega_{0}t + P_{3} \sin 2\omega_{0}t.\label{eq74}
\end{equation}

\noindent We have (on integration)\\
\begin{eqnarray}
X_{1F}&=& -\frac{A^{2}}{2\alpha}(\frac{1}{2} +\frac{b}{\beta}) + \frac{2\omega_{0}P_{3}-\beta P_{2}}{3\omega_{0}^{2}(4\omega_{0}^{2} + \beta^{2})}\cos 2\omega_{0}t\nonumber\\
&-&\frac{\beta P_{3}+ 2\omega_{0}P_{2}}{3\omega_{0}^{2}(4\omega_{0}^{2} + \beta^{2})}\sin 2\omega_{0}t\label{eq75}
\end{eqnarray}
\noindent which, on using Eqs.(\ref{eq70})-(\ref{eq73}) leads to\\
\begin{equation}
X_{1F}=\frac{A^{2}}{2\alpha} \left[-\alpha_{1} + \alpha_{2}\cos 2\omega_{0}t +\alpha_{3} \sin 2\omega_{0}t \right]\label{eq76}
\end{equation}

\noindent where

\begin{eqnarray}
\alpha_{1}&=& \frac{1}{2}+\frac{b}{\beta}\label{eq77}\\
\alpha_{2}&=& \frac{1}{4\omega_{0}^{2} +\beta^{2}} \left[\frac{\beta^{2}}{2} +\frac{1}{3}(1+\sigma)\beta +\frac{4}{3}\omega_{0}^{2}+\frac{2}{3}\omega_{0}^{2}\frac{1+b}{\sigma}\right]\nonumber\\\label{eq78}\\
\alpha_{3}&=&\frac{\omega_{0}}{4\omega_{0}^{2}+{\beta^{2}}}\left[\frac{\beta}{3} +\frac{1}{3}(1+\sigma)-\frac{\beta(1+b)}{3\sigma}\right]\label{eq79}
\end{eqnarray}

\noindent Here, by $X_{1F}$ we mean the finite part of $X_{1}$ coming from the regular terms of Eq.(\ref{eq69}) only. Using this in Eq.(\ref{eq54}), we get the finite part of $Y_{1}$ as\\
\begin{equation}
Y_{1F}=\frac{A^{2}}{2\alpha} \left[-\alpha_{1} + \beta_{2}\cos 2\omega_{0}t + \beta_{3}\sin 2\omega_{0}t \right]\label{eq80}
\end{equation}

\noindent where\\
\begin{eqnarray}
\beta_{2}&=&\alpha_{2} +\frac{2\omega_{0}\alpha_{3}}{\sigma}\label{eq81}\\
\mathrm{~~and~~}\beta_{3}&=&\alpha_{3} -\frac{2\omega_{0}\alpha_{2}}{\sigma}.\label{eq82}
\end{eqnarray}
\noindent Similarly, using Eqs.(\ref{eq76}), (\ref{eq77}) along with Eqs.(\ref{eq65}) and (\ref{eq67}) in Eq.(\ref{eq55}) (this time, considering the nonlinear $XZ$ term) we get\\
\begin{eqnarray}
Z_{1F} &=& \frac{1}{\alpha}\left[X_{1F} -(D+1)Y_{1F} -X_{0}Z_{0}\right]\nonumber\\
&=&\frac{A^{2}}{2\alpha}\left[-\gamma_{1} + \gamma_{2}\cos 2\omega_{0}t +\gamma_{3}\sin 2\omega_{0}t\right]\label{eq83}
\end{eqnarray}
\noindent where\\

\begin{eqnarray}
\gamma_{1} &=&\frac{\omega_{0}^{2}}{\sigma\alpha}\label{eq84}\\
\gamma_{2}&=&-\left[\frac{\omega_{0}^{2}}{\sigma\alpha} + \frac{2\omega_{0}\alpha_{3}}{\sigma\alpha}+\frac{2\omega_{0}}{\alpha}(\alpha_{3}-\frac{2\omega_{0}\alpha_{2}}{\sigma})\right]\label{eq85}\\
\mathrm{~~and~~}\gamma_{3}&=&-\frac{\omega_{0}}{\alpha}(1+\frac{1}{\sigma}) + \frac{2\alpha_{2}\omega_{0}}{\alpha\sigma}+\frac{2\omega_{0}}{\alpha}(\alpha_{2}+\frac{2\alpha_{3}\omega_{0}}{\sigma}).\nonumber\\\label{eq86}
\end{eqnarray}

\noindent This completes our calculation for the finite parts of $X$, $Y$ and $Z$ at the first-order of perturbation.\\
\noindent We now proceed to the second-order by writing the equation\\

\begin{eqnarray}
LX_{2}&=&-\alpha\sigma(X_{0}Y_{1F} +Y_{0}X_{1F})\nonumber\\
&& -\sigma(D+b)(X_{0}Z_{1F}+X_{1F}Z_{0})\nonumber\\
&-&2\sigma b\Delta r A\cos \omega_{0}t +b\omega_{0}\Delta r A\sin\omega_{0}t\label{eq87}
\end{eqnarray}

\noindent from Eq.(\ref{eq59}) above. The last two secular terms on the RHS of Eq.(\ref{eq87}) have been borrowed from Eq.(\ref{eq69}), where we had intentionally suppressed these terms and studied only the finite contributions coming from the regular terms only. The reason for this (explained earlier in the discussion following Eq.(\ref{eq69})) becomes more succinct now. Had we considered these secular terms in the first order then the $\cos \omega_{0} t$ and $\sin \omega_{0}t$ terms coming by integrating them (see Eq.(\ref{eq39}) above) would not have given any new secular terms when producted with $X_{0}$ and $Y_{0}$ in Eq.(\ref{eq87}). Therefore, shifting these secular terms from the first order equation Eq.(\ref{eq69}) to the second order equation Eq.(\ref{eq87}) does not affect the structure of the RG-flow equations evaluated upto the second-order. Our present focus is only on the secular terms in the RHS of Eq.(\ref{eq87}).\\
\noindent Using expressions of $X_{0}$, $Y_{0}$ and $Y_{0}$ [from Eqs.(\ref{eq65}) to (\ref{eq67})] along with those of $X_{1}$, $Y_{1}$ and $Z_{1}$ [from Eqs.(\ref{eq76}),(\ref{eq80}) and (\ref{eq69})] we evaluate the following:\\
\begin{equation}
i)  X_{0}Y_{1F}+X_{1F}Y_{0}\Rightarrow\frac{A^{3}}{2\alpha}\left[\mu_{1} \cos\omega_{0}t+\mu_{2} \sin\omega_{0}t \right]\label{eq88}
\end{equation}

\noindent where `$\Rightarrow$' means `secular terms only' and\\
\begin{eqnarray}
\mu_{1}&=&- 2\alpha_{1} + \frac{\alpha_{2} +\beta_{2}}{2} -\frac{\omega_{0}\alpha_{3}}{\sigma}\label{eq89}\\
\mu_{2}&=&\frac{\alpha_{3}+\beta_{3}}{2} +\frac{\omega_{0}}{\sigma}(\alpha_{1} + \frac{\alpha_{2}}{2})\label{eq90}
\end{eqnarray}

\begin{equation}
ii) X_{0}Z_{1F} +X_{1F}Z_{0}\Rightarrow\frac{A^{3}}{2\alpha}\left[\lambda_{1}\cos \omega_{0}t+\lambda_{2}\sin \omega_{0}t\right]\label{eq91}
\end{equation}

\noindent with\\
\begin{eqnarray}
\lambda_{1}&=& -\gamma_{1} +\frac{\gamma_{2}}{2} +\frac{1}{2\alpha}(\frac{1}{\sigma}+1)\omega_{0}\alpha_{3}-\frac{\omega_{0}^{2}}{\alpha\sigma}(\alpha_{1}-\frac{\alpha_{2}}{2})\nonumber\\\label{eq92}\\
\lambda_{2} &=& \frac{\gamma_{3}}{2}-\frac{\omega_{0}\alpha_{1}}{\alpha}(\frac{1}{\sigma}+1)-\frac{1}{2\alpha}(\frac{1}{\sigma}+1)\omega_{0}\alpha_{2}+\frac{\omega_{0}^{2}\alpha_{3}}{2\sigma\alpha}\nonumber\\\label{eq93}
\end{eqnarray}
\noindent Putting all this back in Eq.(\ref{eq87}) we get the secular terms in the RHS of that equations as,\\
\begin{eqnarray}
LX_{2}&=&-\frac{A^{3}}{2\alpha}C_{1}\cos \omega_{0}t-\frac{A^{3}}{2\alpha}C_{2}\sin \omega_{0}t\nonumber\\
&-& 2\sigma b\Delta r A\cos\omega_{0}t+b\omega_{0}\Delta r A\sin\omega_{0}t\nonumber\\&+&\mathrm{~non-resonant~terms~}\label{eq94}
\end{eqnarray}

\noindent where\\
\begin{eqnarray}
C_{1}&=& \alpha\sigma\mu_{1} + \sigma b \lambda_{1} +\sigma \omega_{0} \lambda_{2}\label{eq95}\\
C_{2}&=& \alpha\sigma\mu_{2} + \sigma b \lambda_{2} -\sigma \omega_{0} \lambda_{1}.\label{eq96}
\end{eqnarray}

\noindent Having done all the necessary calculations, we have cast the RHS of Eq.(\ref{eq94}) is the generic form of Eq.(\ref{eq34}) (see \textbf{Sec.III}). Therefore, now we can directly write down the RG-equation by using the results of \textbf{Sec.III}. We shall have just one RG-equation (for A) here as because in writing the solution of $X_{0}$ in Eq.(\ref{eq65}) we had chosen the appropriate initial conditions accordingly. This, obviously, is not any simplification, but saves cumbersome algebra. Thus, our sought after RG- flow equation for the amplitude $A$ is obtained by combining Eqs.(\ref{eq38}), (\ref{eq39}) and (\ref{eq46}) as,\\
\begin{eqnarray}
\frac{dA}{dt}&=&-\frac{1}{2\omega_{0}(\omega_{0}^{2}+\beta^{2})}\left[\varepsilon\{(-2\sigma b \Delta r A\omega_{0}+b\omega_{0}\Delta r A \beta)\}\right.\nonumber\\
&&+\left.\varepsilon^{2}\{-\frac{A^{3}}{2\alpha}C_{1}\omega_{0}-\frac{A^{3}}{2\alpha}C_{2}\beta\}\right]\nonumber\\
&=&\frac{A}{2(\omega_{0}^{2}+\beta^{2})}\left[\varepsilon(\sigma-b-1)b \Delta r \right.\nonumber\\
&& \left.+\frac{\varepsilon^{2}}{2\alpha}(C_{1} + \frac{C_{2}\beta}{\omega_{0}})A^{2}\right] \label{eq97}
\end{eqnarray}

\noindent where we have used the form of the operator $L$ as in Eq.(\ref{eq62}) with $\beta=\sigma+b+1$ (compare this with Eq.(\ref{eq35})). Also, we have associated the terms on the RHS of Eq.(\ref{eq97}) with the appropriate powers of $\varepsilon$, to lay bare the orders of perturbation from which they have come. The structure of the above amplitude equation [Eq.(\ref{eq97})] is reminiscent of the general discussions on Hopf bifurcations we made at the beginning of \textbf{Sec:II} (see the discussions following Eq.(\ref{eq2})). In those lines, it is clear from the RHS of Eq.(\ref{eq97}) that Hopf bifurcation occurs right at the point where the co-efficient of the linear term (in A) vanishes i.e. when \\
\begin{equation}
\Delta r=0\Rightarrow r=r_{0} \mathrm{~(Hopf~point~)~}.\label{eq98}
\end{equation}

\noindent This ensures that the $A^{3}$ term comes to the forefront as the only player to lead the system towards a super-critical or a sub-critical Hopf bifurcation, depending on whether its co-efficient is negative or positive respectively. To understand these bifurcations, we first go to one of the extremes and take $\sigma$ very large. Then from Eqs.(\ref{eq60}),(\ref{eq69}) and (\ref{eq70}) we have,\\
\begin{equation}
r_{0}\approx \sigma,~\omega_{0}^{2} \approx 2b\sigma \mathrm{~and~} \beta\approx\sigma.\label{eq99}
\end{equation}

\noindent Accordingly from Eqs.(\ref{eq77})-(\ref{eq79}) we get\\

\begin{eqnarray}
\alpha_{1} &=&\frac{1}{2} +\frac{b}{\beta}\approx\frac{1}{2} + \frac{b}{\sigma}\approx\frac{1}{2}\label{eq100}\\
\alpha_{2}&\approx& \frac{1}{(8b\sigma+\sigma^{2})}\left[\frac{\sigma^{2}}{2}+\frac{1}{3}(1+\sigma)\sigma+\frac{8}{3}b\sigma+\frac{4}{3}b\sigma \frac{1+b}{\sigma}\right]\nonumber\\
&\approx& \frac{1}{\sigma^{2}}\left[\frac{\sigma^{2}}{2}+\frac{\sigma^{2}}{3}\right]=\frac{5}{6}\label{101}\\
\alpha_{3}&=&\frac{\omega_{0}}{8b\sigma+\sigma^{2}}\left[\frac{\sigma}{3}+\frac{2}{3}(1+\sigma)+\frac{1+b}{3}\right]\approx\frac{\omega_{0}}{\sigma^{2}}(\frac{\sigma}{3}+\frac{2}{3}\sigma)\nonumber\\
&=&\frac{\omega_{0}}{\sigma}.\label{102}
\end{eqnarray}

\noindent Similar approximations (with $\sigma\rightarrow\infty$) in Eqs.(\ref{eq81}),(\ref{eq82}),(\ref{eq84})-(\ref{eq86}),(\ref{eq89}),(\ref{eq90}),(\ref{eq92}),(\ref{eq93}) leads to\\

\begin{eqnarray}
\beta_{2}&\approx& \frac{5}{6}, \beta_{3}\approx -\frac{2}{3}\frac{\omega_{0}}{\sigma}\nonumber\\
\gamma_{1}&\approx&\frac{2b}{\alpha},\gamma_{2}\approx\frac{2}{3}\frac{b}{\alpha}, \gamma_{3}\approx\frac{2}{3}\frac{\omega_{0}}{\alpha}\nonumber\\
\mu_{1}&\approx& -\frac{1}{6}, \mu_{2}\approx\frac{13}{12}\frac{\omega_{0}}{\sigma}\nonumber\\
\lambda_{1}&\approx&-\frac{5}{6}\frac{b}{\alpha}, \lambda_{2}\approx -\frac{7}{12}\frac{\omega_{0}}{\alpha}.\label{eq103}
\end{eqnarray}

\noindent Putting Eqs.(\ref{eq100})-(\ref{eq103}) into the Eqs.(\ref{eq95}) and (\ref{eq96}) we get finally the sign of the co-efficient of $A^{3}$ in the amplitude equation (Eq.(\ref{eq97})) from the sign of\\
\begin{eqnarray}
C_{1} + C_{2}\frac{\beta}{\omega_{0}}&\approx& -\frac{7}{6}\frac{b}{\alpha}\sigma^{2}+\frac{1}{4}\frac{\sigma\omega_{0}b}{\alpha}\frac{\sigma}{\omega_{0}}\nonumber\\
&=&-\frac{11}{12}\frac{b}{\alpha}\sigma^{2}<0\label{eq104}
\end{eqnarray}

\noindent which is negative, vindicating the fact that for very large $\sigma$ we have ($r=r_{0}$) as the point of supercritical Hopf bifurcation. But that is not always the case. For a moderate value of the Prandtl number as $\sigma=10$ and the parameter $b=\frac{8}{3}$ [these are precisely the values that Lorenz used in his original simulation \cite{stro}] we get from Eq.(\ref{eq60}) the values $r_{0}=24.74$, $\omega_{0}^{2}=92.63$ and accordingly the following set of values for the various constants follow:-\\
\begin{eqnarray}
\alpha_{1}&=&0.7, \alpha_{2}=0.5, \alpha_{3}=0.18\nonumber\\
\beta_{2}&=&1.0, \beta_{3}=-0.7\nonumber\\
\gamma_{1}&=&1.18, \gamma_{2}=0.46, \gamma_{3}=1.6\nonumber\\
\mu_{1}&=&-0.74,\mu_{2}=0.65\nonumber\\
\lambda_{1}&=&-1.36, \lambda_{2}=-0.4.\label{eq105}
\end{eqnarray}

\noindent All these lead to $C_{1}=-133.66$ and $C_{2}=171.96$ and hence the sign of the co-efficient of $A^{3}$
in Eq.(\ref{eq97}) is obtained from the sign of \\
\begin{equation}
C_{1}+C_{2}\frac{\beta}{\omega_{0}}=110.52>0\label{eq106}
\end{equation}

\noindent which is positive thus signalling at a subcritical Hopf bifurcation. Therefore, there is a critical value of the Prandtl number ($\sigma$) below which the Hopf bifurcation is subcritical and above which supercritical. This is the information we extract from the amplitude equation Eqs.(\ref{eq97}) derived using RG.\\

\noindent The results that we have obtained are in agreement with all available numerical data and a specific $\sigma =10$ calculation of \cite{mm}. For $\sigma=10$,[Fig 3] the Hopf bifurcation is known to be subcritical and for $\sigma=50$, \cite{Fow} found a periodic orbit for $r>r_{0}$.

\begin{figure}
{\includegraphics[width=0.50\textwidth,angle=0,clip]{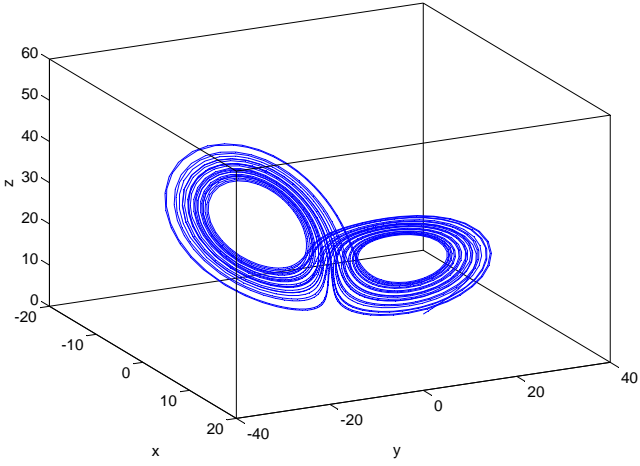}}
\caption[]{Numerical phase plot for Lorenz attractor. Here $\sigma=10$ and $b=\frac{8}{3}$.}
\end{figure}

\noindent We close this Section by mentioning that there is a lower bound on the Prandtl number $\sigma$, for the above analysis to make sense. That is obtained trivially as $\sigma>b+1$ from the definitions of $r_{0}$ and $\omega_{0}$ in Eqs.(\ref{eq60}) and (\ref{eq61}) and the requirement that all parameters of the Lorenz system are positive, failing which, the fixed points $x_{0}=y_{0}=\pm\sqrt{b(r-1)}$ and $z_{0}=r-1$ are stable for all $r>1$ thus obscuring any Hopf bifurcation.

\section*{V:ROSSLER MODEL}
In this Section we study plausible Hopf bifurcations in the Rossler model. Having detailed the methodology and its application for the Lorenz model in the last two sections, our discussion in this Section will be brief as because the main line of approach remains the same. The Rossler systems is given by the three equations\\
\begin{eqnarray}
Dx&=&-y-z \label{eq107}\\
Dy&=&x+ay\label{eq108}\\
Dz&=&b+xz-cz\label{eq109}
\end{eqnarray}
\noindent where $D\equiv\frac{d}{dt}$ and $a,b,c$ are adjustable parameters. These three equations can be combined to give a single variable equation in $y$ as\\

\begin{eqnarray}
&&D^{3}y + (c-k)D^{2}y + (1+kc)Dy +cy \nonumber\\
&=&\dot{y}\ddot{y}-k\dot{y}^{2} +k^{2}y\dot{y} + y\dot{y}-ky^{2}-k\label{eq110}
\end{eqnarray}
\noindent where $a=b=k$ has been used. This is not any essential restriction and numerous numerical experiments can be carried out with various values of $a$, $b$ and $c$. But for the sake of algebraic simplicity we stick to equal values of $a$ and $b$ here and focus on Hopf bifurcations as the parameter $c$ is varied.\\
\noindent The fixed points of the Rossler system Eqs.(\ref{eq107})-(\ref{eq109}) are obtained at\\
\begin{eqnarray}
x_{0}&=&z_{0}=\frac{d}{2}\label{eq111}\\
y_{0}&=&-\frac{d}{2a}\label{eq112}
\end{eqnarray}

\noindent with\\
\begin{equation}
d=c\pm\sqrt{c^{2}-4ab}.\label{eq113}
\end{equation}

\noindent Shifting our origin to one of the fixed points as $u_{1}=x-x_{0}$, $u_{2}=y-y_{0}$ and $u_{3}=z-z_{0}$, we can recast Eq.(\ref{eq110}) as\\
\begin{eqnarray}
&&\dddot{u}+(c-k+ky_{0})\ddot{u}+(1-kc-k^{2}y_{0}-y_{0})\dot{u}\nonumber\\
&+&(c+2ky_{0})u\nonumber\\
&=&\dot{u}\ddot{u}-ku\ddot{u}-k\dot{u}^{2}+(k^{2}+1)u\dot{u}-ku^{2}\label{eq114}
\end{eqnarray}

\noindent where to simplify notation, `$u$' has been written in place of $u_{2}$ in the above equation. From this point our focus will be to derive an amplitude equation where a quantity like\\
\begin{equation}
\Delta c=c-c_{0}\label{eq115}
\end{equation}

\noindent appears whose zero value corresponds to the Hopf-bifurcation point (this role was played by the parameter `$r$' the Lorenz system). If there is a Hopf bifurcation in the Rossler system, then we expect $\Delta c$ to appear in the co-efficient of the linear term (in amplitude) in the amplitude equation. Only then can we infer that a Hopf bifurcation occurs at $c=c_{0}$.\\
\noindent Expanding $y_{0}$ in Eq.(\ref{eq112}) about $c_{0}$ we get\\

\begin{equation}
y_{0}=\alpha_{1}+\alpha_{2} \Delta c\label{eq116}
\end{equation}
\noindent where\\

\begin{eqnarray}
\alpha_{1}&=&\frac{-c_{0}\pm \sqrt{c_{0}^{2}-4ab}}{2a}\label{eq117}\\
\alpha_{2}&=&-\frac{1}{2a}\pm \frac{c_{0}}{2a\sqrt{c_{0}^{2}-4ab}}.\label{eq118}
\end{eqnarray}

\noindent Inserting these in Eq.(\ref{eq114}), allows us to write that equation as,\\
\begin{eqnarray}
Lu&=&\frac{1}{2}\frac{d}{dt}(\dot{u}^{2})-k\frac{d}{dt}(u\dot{u})+\frac{1+k^{2}}{2}\frac{d}{dt}(u^{2})-ku^{2}\nonumber\\
&-&\Delta c \{(1+\alpha_{2})\ddot{u}-(k+(1+k^{2})\alpha_{2})\dot{u}+(1+2k\alpha_{2})u\}\nonumber\\\label{eq119}
\end{eqnarray}

\noindent where the operator $L$ is product separable as \\
\begin{equation}
Lu=(D^{2}+\omega_{0}^{2})(D+\sigma)u\label{eq120}
\end{equation}
\noindent with $\omega_{0}$ and $\sigma$ given by
\begin{eqnarray}
\omega_{0}^{2}&=&\frac{c_{0}-2x_{0}}{c_{0}-x_{0}-k}\label{eq121}\\
\sigma &=& c_{0}-x_{0}-k\label{eq122}
\end{eqnarray}

\noindent These values of $\omega_{0}$ and $\sigma$ are easily obtainable by comparing the cubic operator on the RHS of Eq.(\ref{eq120}) with the cubic characteristic-equation of the linearized Rossler model.\\
\noindent Now, as was done in studying the Lorenz model, we invoke the perturbation expansion in $u$ as\\
\begin{equation}
u=u_{0}+\lambda u_{1} +\lambda^{2}u_{2}\label{eq123}
\end{equation}

\noindent with $\lambda$ as the perturbation parameter. Using this expansion in Eq.(\ref{eq119}) we can easily segregate the terms of different orders and obtain the equations for zeroth and first orders as \\
\begin{equation}
Lu_{0}=0\label{eq124}
\end{equation}
\begin{eqnarray}
Lu_{1}&=&\beta_{1}+\beta_{2}\sin2\omega_{0}t+\beta_{3}\cos2\omega_{0}t\nonumber\\
&+&\Delta c(\gamma_{1}\cos \omega_{0}t+\gamma_{2}\sin\omega_{0}t)\label{eq125}
\end{eqnarray}

\noindent where we have used the following abbreviations:\\
\begin{eqnarray}
\beta_{1}&=&-\frac{kA^{2}}{2},~\beta_{2}=\frac{1}{2}\omega_{0}^{3}A^{2}-\frac{1}{2}(1+k^{2})\omega_{0}A^{2}\nonumber\\
\beta_{3}&=&k\omega_{0}^{2}A^{2}-\frac{k}{2}A^{2}\nonumber\\
\gamma_{1}&=&A\omega_{0}^{2}(1+\alpha_{2})-A(1+2\alpha_{2})\nonumber\\
\gamma_{2}&=&-\left[k+(1+k^{2})\alpha_{2}\right]A\omega_{0}.\label{eq126}
\end{eqnarray}

\noindent Emulating our approach for the Lorenz model, we write the solution of Eq.(\ref{eq124}) as\\
\begin{equation}
u_{0}=A\cos\omega_{0}t.\label{eq127}
\end{equation}

\noindent For the first order equation, the solution for the regular (non-resonant) part is \\
\begin{equation}
u_{0}=u_{1F}=\delta_{1}+\delta_{2}\sin2\omega_{0}t+\delta_{3}\cos2\omega_{0}t\label{eq128}
\end{equation}
\noindent where $u_{1F}$ represents the finite (non-divergent) part of $u_{1}$ and\\
\begin{eqnarray}
\delta_{1}&=&\frac{\beta_{1}}{\sigma},~\delta_{2}=\frac{\beta_{2}\sigma+2\beta_{3}\omega_{0}}{4\omega_{0}^{2}+\sigma^{2}}\nonumber\\
\delta_{3}&=&\frac{\beta_{3}\sigma-2\beta_{2}\omega_{0}}{4\omega_{0}^{2}+\sigma^{2}}.\label{eq129}
\end{eqnarray}

\noindent The divergent (resonant) terms on the RHS of Eq.(\ref{eq125}) can be (as was done in the Lorenz model) stacked with the divergent terms of the second order equation,\\
\begin{eqnarray}
Lu_{2}&=&\frac{1}{2}\frac{d}{dt}(\dot{u}_{0}\dot{u}_{1F})-k\frac{d}{dt}(u_{0}\dot{u}_{1F}+u_{1F}\dot{u}_{0})\nonumber\\
&+&\frac{1+k^{2}}{2}\frac{d}{dt}(2u_{0}u_{1F})-2ku_{0}u_{1F}\nonumber\\
&+&\Delta c(\gamma_{1}\cos \omega_{0}t+\gamma_{2}\sin\omega_{0}t)\label{eq130}
\end{eqnarray}
\noindent because (see explanation following Eq.(\ref{eq87})), had we integrated the resonant terms in the first order, then the regular $\cos\omega_{0}t$ and $\sin\omega_{0}t$ terms coming from there, would not have product any new secular terms in the different produced appearing on the RHS of Eq.(\ref{eq130}). \\
\noindent Identifying the secular terms from the RHS of Eq.(\ref{eq130}) we find,\\
\begin{eqnarray}
Lu_{2}&=&A^{3}C_{1}\cos\omega_{0}t+A^{3}C_{2}\sin\omega_{0}t\nonumber\\
&+&A \Delta c \gamma_{1}\cos\omega_{0}t + A \Delta c\gamma_{2} \sin\omega_{0}t\nonumber\\
&+&\mathrm{~regular~terms~}\label{eq131}
\end{eqnarray}

\noindent where\\
\begin{eqnarray}
C_{1}&=&\frac{1}{A^{2}}\left[\left(\frac{\omega_{0}^{3}}{2}+\frac{1+k^{2}}{2}\omega_{0}\right)\delta_{2}\right.\nonumber\\
&+&\left. \left(\frac{\omega_{0}^{2}k}{2}-k\right)\delta_{3}+(\omega_{0}^{2}k-2k)\delta_{1}\right]\nonumber\\
C_{2}&=&\frac{1}{A^{2}}\left[-\left(\frac{\omega_{0}^{3}}{2}+\frac{1+k^{2}}{2}\omega_{0}\right)\delta_{3}+(\frac{\omega_{0}^{2}k}{2}-k)\delta_{2}\right.\nonumber\\
&-&\left.\frac{1+k^{2}}{2}\omega_{0}\delta_{1}\right].\label{eq132}
\end{eqnarray}

\noindent The solution of Eq.(\ref{eq131}) is obtained as \\
\begin{eqnarray}
u_{2}&=&-\frac{A.\Delta c.(\gamma_{1}\omega_{0}+\gamma_{2}\sigma)+A^{3}(C_{1}\omega_{0}+C_{2}\sigma)}{2\omega_{0}(\omega_{0}^{2}+\sigma^{2})}t\cos\omega_{0}t\nonumber\\
&+&\frac{A.\Delta c.(\gamma_{1}\sigma-\gamma_{2}\omega_{0})+A^{3}(C_{1}\sigma-C_{2}\omega_{0})}{2\omega_{0}(\omega_{0}^{2}+\sigma^{2})}t\sin\omega_{0}t\nonumber\\
&+&\mathrm{regular ~part}.\label{eq133}
\end{eqnarray}

\noindent Going by the methodology developed in \textbf{Sec.III} we obtain the RG-equation for the amplitude by combining Eqs.(\ref{eq38}), (\ref{eq39}) and (\ref{eq46}) as\\

\begin{eqnarray}
\frac{dA}{dt}&=&-\frac{A}{2\omega_{0}}\frac{1}{\omega_{0}^{2}+\sigma} \left [\lambda(\gamma_{1}\omega_{0}+\gamma_{2}\sigma)\Delta c \right.\nonumber\\
&&\left.+\lambda^{2}(C_{1}+C_{2}\frac{\sigma}{\omega_{0}})A^{2}\right].\label{eq134}
\end{eqnarray}

\noindent This bears resemblance with Eq.(\ref{eq97}), (i.e., the amplitude equation for the Lorenz model) in that, the co-efficient of the linear term `$A$' has a $\Delta c$, which becomes zero at the Hopf-point. Thus,\\
\begin{equation}
c=c_{0}\mathrm{~~(Hopf~point)~}\label{eq135}
\end{equation}

\noindent is the point in parameter space where the system undergoes Hopf bifurcation.\\
\noindent To illustrate, we consider two distinct points in parameter space,\\
\noindent i) $a_{0}=b_{0}=0.2, c_{0}=5.7$ [see Fig.4]\\
\noindent ii)$a_{0}=b_{0}=0.1, c_{0}=14.0$[see Fig.5]\\
\noindent These values are well-known in numerical experiments done with the Rossler systems in context of Hopf bifurcations \cite{stro}. For case (i), we have the values from Eqs.(\ref{eq118}), (\ref{eq121}), (\ref{eq122}), (\ref{eq126}), (\ref{eq132}) as:-\\

\begin{eqnarray}
a&=&b=0.2;~c=5.7\nonumber\\
\omega_{0}&=&5.43;~ \alpha_{2}=-5.0062;~ \sigma=-0.193\nonumber\\
\beta_{1}&=&-0.1A^{2};~\beta_{2}=77.1276A^{2};~\beta_{3}=5.7922A^{2}\nonumber\\
\delta_{1}&=&0.5181A^{2};~\delta_{2}=0.4071A^{2};~ \delta_{3}=-7.1121\nonumber\\
\gamma_{1}&=&-109.014A;~\gamma_{2}=27.1739A\nonumber\\
C_{1}&=&-2.9664;~ C_{2}=588.3718.\label{eq136}
\end{eqnarray}

\begin{figure}
{\includegraphics[width=0.50\textwidth,angle=0,clip]{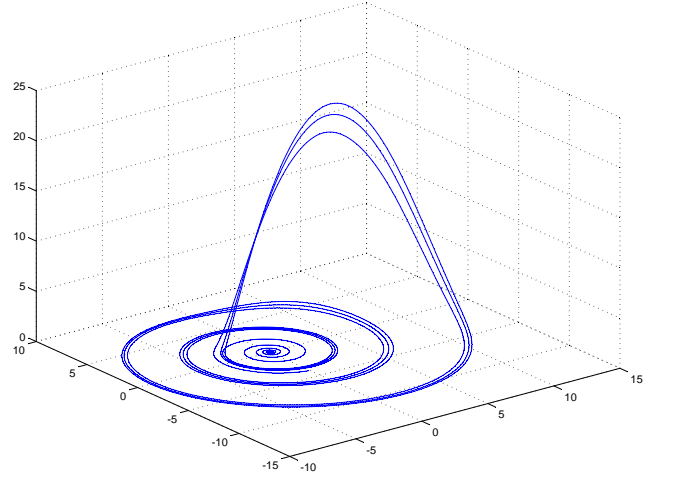}}
\caption[]{Numerical phase plot for Rossler attractor. Here $a = b = 0.2$ and $c = 5.7$.}
\end{figure}

\noindent Putting these values in the amplitude equation we obtain\\
\begin{equation}
\frac{dA}{dt}=\frac{1}{22(\omega_{0}^{2}+\sigma^{2})}(23.89)A^{3}\label{eq137}
\end{equation}

\noindent at the Hopf-bifurcation point. Here the coefficient of $A^{3}$ being positive we understand that a subcritical Hopf bifurcation occurs at $a_{0}=b_{0}=0.2$ and $c_{0}=5.7$.\\
\noindent For case (ii) on the other hand, we obtain the following set of values:-\\

\begin{figure}
{\includegraphics[width=0.50\textwidth,angle=0,clip]{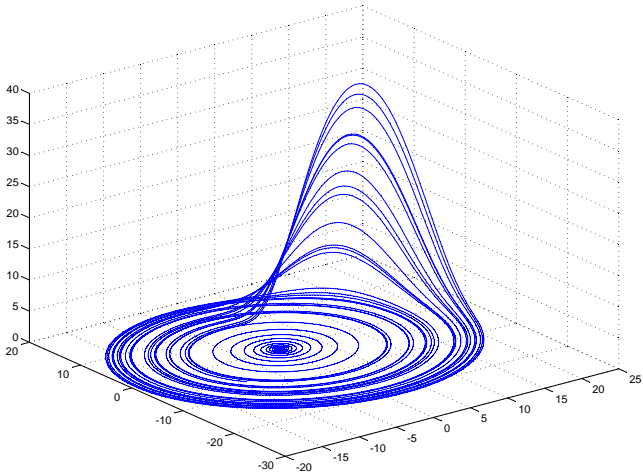}}
\caption[]{Numerical phase plot for Rossler attractor. Here $a = b = 0.1$ and $c = 14.0$.}
\end{figure}

\begin{eqnarray}
a&=&b=0.1;~c=14\nonumber\\
\omega_{0}^{2}&=&140\Rightarrow\omega_{0}=\pm11.83;~ \alpha_{2}=-10\nonumber\\
\beta_{1}&=&-0.05A^{2};~ \beta_{2}=821.83A^{2};~ \beta_{3}=13.95A^{2}\nonumber\\
\delta_{1}&=&0.5A^{2};~\delta_{2}=0.443A^{2};~\delta_{3}=-3.46A^{2}\nonumber\\
\gamma_{1}&=&-1241A;~\gamma_{2}=118.3A\nonumber\\
C_{1}&=&386.321;~C_{2}=2880.55 \label{eq138}
\end{eqnarray}

\noindent which yield the amplitude equation (at the Hopf-point)\\
\begin{equation}
\frac{dA}{dt}=-1.3A^{3}.\label{eq139}
\end{equation}

\noindent This co-efficient being negative, we understand that a super-critical Hopf bifurcation occurs at the point $a_{0}=b_{0}=0.1$ and $c_{0}=14$ of the point parameter space.\\

\section*{VI:SUMMARY}
In this paper the criteria for occurrences of super-critical and sub-critical Hopf bifurcations have been studied for dynamical systems in two as well as three dimensions. In doing so we have employed the renormalization group for perturbatively deriving the corresponding amplitude equation upto some relevant order of the amplitude, where, putting off the linear term in amplitude, the sign of the coefficient of the lowest nonlinear power guides us correctly to the kind of Hopf bifurcation the system shows. This strategy has been successfully applied to first rederive the well known criterion (for two dimensions) which tells one type of Hopf bifurcation from the other and then show some limitations of that criterion through examples where the RG works authentically. The extension of this RG formalism to three dimensions, although nontrivial, has been done and applied to the highly important models of Lorenz and Rossler. The emphasis of the study for these systems has been laid on identifying regions in parameter space where super- or sub-critical Hopf bifurcations can occur. Calculations to second order in perturbation have been done elaborately.\\

\renewcommand{\theequation}{A.\arabic{equation}}
\setcounter{equation}{0}
\section*{APPENDIX : DERIVATION OF Eq.(\ref{eq24})}

\noindent In this Appendix we show how to derive Eq.(\ref{eq19}) and Eq.(\ref{eq23}) using Eq.(\ref{eq15}) and Eq.(\ref{eq16}). At first we analyze the terms on the RHS of Eq.(\ref{eq16}) one by one. The `$-g(x_{0},y_{0})$' series (see Eq.(\ref{eq6})) gives $a^{3}\sin \phi$ from two terms. They are $x^{2}_{0}y_{0}$ and $y_{0}^{3}$, as is obvious from Eqs.(\ref{eq17}), (\ref{eq18}). The former one yields $a^{2}\cos^{2}\phi.a\sin\phi \Rightarrow \frac{a^{3}}{4}\sin\phi$ as the relevant part. These terms give secular divergence as has been discussed earlier (see Eqs.(\ref{eq11}) and (\ref{eq12}) ). Now for the co-efficients $g_{ij}$. In the format of Eq.(\ref{eq6}) the co-efficient of $x^{2}_{0}y_{0}$ is $g_{21}$ and of $y^{3}_{0}$ is $g_{03}$. These two co-efficients can be sieved out of the series Eq.(\ref{eq6}) by taking the Taylor-derivatives as\\
\begin{equation}
g_{21}=\frac{1}{2}g_{xxy}\label{eqA.1}
\end{equation}
\noindent and\\
\begin{equation}
g_{03}=\frac{1}{6}g_{yyy}\label{eqA.2}
\end{equation}
\noindent where the subscripts mean partial derivatives at ($x=0$) and ($y=0$). Therefore, the co-efficient of the secular $sin \phi$ terms coming from the term `$-g(x_{0},y_{0})$' of Eq.(\ref{eq15}) is \\
\begin{equation}
-g(x_{0},y_{0})\Rightarrow-\frac{a^{3}}{8}(g_{xxy} + g_{yyy})\label{eqA.3}
\end{equation}
\noindent where the `$\Rightarrow$' means `relevant secular contribution'.\\
\noindent The other two terms of Eq.(\ref{eq15}) can be similarly analyzed. The second term on the RHS of Eq.(\ref{eq15}) is \\
\begin{equation}
-y_{0}\frac{\partial f}{\partial x}(x_{0},y_{0})=-\Sigma_{i,j} i f_{ij} x_{0}^{i-1}y_{0}^{j+1}.\label{eqA.4}
\end{equation}
\noindent Here the relevant $x^{2}_{0}y_{0}$ and $y_{0}^{3}$ terms are obtained from the combinations $(i=3,j=0)$ and $(i=1,j=2)$ respectively. For the former combination we have $-\frac{3}{4}a^{3}f_{30}=-\frac{1}{8}a^{3}f_{xxx}$ as the relevant secular contribution and for the latter combination we have
$-\frac{3}{4}a^{3}f_{12} = -\frac{3}{8}a^{3}f_{xyy}$ as the relevant secular contribution, thus allowing us to write \begin{equation}
-y_{0}\frac{\partial f}{\partial x}(x_{0},y_{0}) \Rightarrow -\frac{a^{3}}{8}(f_{xxx}+3f_{xyy}).\label{eqA.5}
\end{equation}
\noindent Similarly, for the third term on the RHS of Eq.(\ref{eq15}) we have, \\
\begin{equation}
x_{0}\frac{\partial f}{\partial y} = \Sigma_{i,j} j f_{ij} x_{0}^{i+1}y_{0}^{j-1}\label{eqA.6}
\end{equation}
\noindent which gives only a $x_{0}^{2}y_{0}$ term ($i=1,j=2$) as the relevant one for our purpose, but no $y_{0}^{3}$ term (as $i\geq2$). This term has $2.f_{12}\frac{a^{3}}{4}$ as the co-efficient of $\sin\phi$ which hence leads to\\
\begin{equation}
x_{0}\frac{\partial f}{\partial y}\Rightarrow \frac{a^{3}}{4} f_{xyy}.\label{eqA.7}
\end{equation}
\noindent Adding up Eqs.(\ref{eqA.3}),(\ref{eqA.5}),(\ref{eqA.7}) we have the co-efficient of $\sin(t+\theta)=\sin\phi$ from the first order terms of Eq.(\ref{eq19}).\\

\noindent Now the regular $a^{2}$-terms on the RHS of Eq.(\ref{eq15}) are as follows:-\\
\begin{eqnarray}
-g(x_{0},y_{0})\rightarrow &-&g_{20}x_{0}^{2} - g_{11}x_{0}y_{0}-g_{02}y_{0}^{2}-.....\nonumber\\
&=& -\frac{a^{2}}{2}\left[(g_{20}+g_{02}) + (g_{20}-g_{02})\cos2\phi\right.\nonumber\\
&+&\left.g_{11}\sin2\phi\right]\label{eqA.8}
\end{eqnarray}\\

\begin{equation}
-y_{0}\frac{\partial f}{\partial x}(x_{0},y_{0})\rightarrow-\frac{a^{2}}{2}\left[f_{11}-f_{11}\cos2\phi + 2f_{20}\sin2\phi\right]\label{eqA.9}
\end{equation}
\begin{equation}
x_{0}\frac{\partial f}{\partial y}(x_{0},y_{0})\rightarrow \frac{a^{2}}{2} \left[f_{11} + f_{11}\cos2\phi + 2f_{02}\sin2\phi\right]\label{eqA.10}
\end{equation}
\noindent where `$\rightarrow$' means `relevant regular term'. Adding up the terms we get the regular $a^{2}$-terms on the RHS of Eq.(\ref{eq15}) as,
\begin{eqnarray}
&&\ddot{x_{1}}+x_{1}\nonumber\\
&=&\frac{a^{2}}{4}\left[-(g_{xx} + g_{yy}) + (4f_{xy}- g_{xx} + g_{yy})\cos 2\phi \right.\nonumber\\
&+&\left.2(-f_{xx} +f_{yy} -g_{xy})\sin 2\phi\right]\nonumber\\
&+& \mathrm{~higher~powers~of~}a \label{eqA.11}
\end{eqnarray}

\noindent which, on integration, gives the renormalized $x_{1}$ of Eq.(\ref{eq21}).\\

\noindent Now we turn to Eq.(\ref{eq16}) for the $a^{3}\sin\phi$ terms that come in the second-order of perturbation. There are three types of terms on the RHS of Eq.(\ref{eq16}): \textbf{(i)} Product of $x_{1}$ (or $y_{1}$) with the first derivatives of $f(x,y)$ or $g(x,y)$. \textbf{(ii)} Product of $x_{0}$ (or $y_{0}$), $x_{1}$ (or $y_{1}$) and the second derivatives of $f$ or $g$. \textbf{(iii)} Product of the functions and their first derivatives. Calculations are easy and, in order to illustrate, we pick up one term from each of the above three categories.\\
\noindent Among the four terms of type \textbf{(i)}, viz, $-x_{1}\frac{\partial g}{\partial x}(x_{0},y_{0})$, $x_{1}\frac{\partial f}{\partial y}(x_{0},y_{0})$, $-y_{1}\frac{\partial g}{\partial y}(x_{0},y_{0})$ and $-y_{1}\frac{\partial f}{\partial x}(x_{0},y_{0})$, we work out the case of $-x_{1}\frac{\partial g}{\partial x}$ here, and state the results for the other three terms. For the term $-x_{1}\frac{\partial g}{\partial x}$, it is clear that the $a^{2}$-terms from $x_{1}$ and the linear terms from $\frac{\partial g}{\partial x}$ combine to give the required $a^{3}$-terms. Among these $a^{3}$-terms, the constant and $\cos2\phi$ terms from $x_{1}$ [see Eq.(\ref{eq21})] combine with $y_{0}(=a\sin\phi)$ of $\frac{\partial g}{\partial x}(x_{0},y_{0})$ to give our sought after secular term $a^{3}\sin\phi$. Also, the $\sin2\phi$ term of $x_{1}$ combine with $x_{0}(=a\cos\phi)$ term of $\frac{\partial g}{\partial x}(x_{0},y_{0})$ to give the same.\\
\noindent Thus considering only $a^{3}$-terms, we have, \\
\begin{eqnarray}
-x_{1}\frac{\partial g}{\partial x}(x_{0},y_{0}) &\Rightarrow& -x_{1} \left[2g_{20}x_{0} + g_{11}y_{0}\right]\nonumber\\
&=& -x_{1}\left[g_{xx}x_{0} + g_{xy}y_{0}\right]\label{eqA.12}
\end{eqnarray}
\noindent which, on using Eqs.(\ref{eq17}),(\ref{eq18}),(\ref{eq21}) leads to the expression\\
\begin{eqnarray}
\Rightarrow &&a^{3}\sin\phi \left[\frac{1}{12} (f_{yy} -f_{xx})g_{xx} + \frac{5}{24}(g_{xx} + g_{yy})g_{xy}\right.\nonumber\\
&&-\left.\frac{1}{6}f_{xy}g_{xy}\right]\label{eqA.13}
\end{eqnarray}

\noindent For the other three terms of type \textbf{(i)} we have:-\\
\begin{eqnarray}
x_{1}\frac{\partial f}{\partial y}(x_{0},y_{0})&\Rightarrow& x_{1}\left[f_{11} x_{0}+2 f_{02}y_{0}\right]\nonumber\\
&\Rightarrow& a^{3}\sin\phi \left[\frac{1}{12}f_{xy}(f_{xx} + f_{yy} +g_{xy})\right.\nonumber\\
&&-\left.\frac{1}{24}f_{yy}(7g_{xx} + 5g_{yy})\right]\label{eqA.14}
\end{eqnarray}
\begin{eqnarray}
&&-y_{1}\frac{\partial g}{\partial y}(x_{0},y_{0})\Rightarrow -y_{1}\left[g_{11}x_{0} + 2g_{02}y_{0} \right]\nonumber\\
&\Rightarrow& a^{3} \sin\phi\left[\frac{1}{12}g_{xy}(f_{xy}-g_{xx}-g_{yy})\right.\nonumber\\
&-& \left.\frac{1}{24} g_{yy}(7f_{xx} + 5f_{yy})\right]\label{eqA.15}
\end{eqnarray}
\noindent and\\
\begin{eqnarray}
&&-y_{1}\frac{\partial f}{\partial x}(x_{0},y_{0})\Rightarrow -y_{1}\left[2f_{20}x_{0} + f_{11} y_{0}\right]\nonumber\\
&\Rightarrow& a^{3}\sin\phi \left[\frac{1}{12}(g_{yy}-g_{xx})f_{xx}-\frac{5}{24}(f_{xx} + f_{yy})f_{xy}\right.\nonumber\\
&-&\left.\frac{1}{6}f_{xy}g_{xy}\right].\label{eqA.16}
\end{eqnarray}

\noindent Adding up Eq.(\ref{eqA.12}) to Eq.(\ref{eqA.16}), we get the relevant secular contribution from the type-\textbf{(i)} terms as,\\
\begin{eqnarray}
&-&x_{1} \frac{\partial g}{\partial x}(x_{0},y_{0}) \nonumber\\
&+& x_{1} \frac{\partial f}{\partial y}(x_{0},y_{0}) -y_{1}\frac{\partial g}{\partial y}(x_{0},y_{0})-y_{1}\frac{\partial f}{\partial x}(x_{0},y_{0})\nonumber\\
&\Rightarrow& a^{3} \sin\phi \left[-\frac{1}{6}(f_{xx}g_{xx} + f_{xy}g_{xy})\right.\nonumber\\ &-&\left.\frac{5}{24}(f_{yy}g_{xx}+2f_{yy}g_{yy}+f_{xx}g_{yy}) + \frac{1}{8}g_{xy}(g_{xx}+g_{xy})\right.\nonumber\\
&-& \left. \frac{1}{8} f_{xy}(f_{xx} + f_{xy})\right].\nonumber\\\label{eqA.17}
\end{eqnarray}
\noindent The type -\textbf{(ii)} terms of Eq.(\ref{eq16}) are $-y_{0} x_{1}$ $\frac{\partial^{2} f}{\partial x^{2}}(x_{0},y_{0})$, $-y_{0}y_{1}\frac{\partial^{2} f}{\partial x \partial y}(x_{0},y_{0})$, $-x_{0}x_{1}\frac{\partial^{2}f}{\partial x\partial y}(x_{0},y_{0})$ and $x_{0}y_{1}$ $\frac{\partial^{2} f}{\partial y^{2}}(x_{0},y_{0})$, of which we elucidate only the first one. These terms are simpler than those of type \textbf{(i)}. In the term $-y_{0}x_{1}\frac{\partial^{2}f}{\partial x^{2}}(x_{0},y_{0})$, the $[y_{0}x_{1}]$ part gives $a^{3}$.\\
\noindent It is only the $f_{20}$ term (co-efficient of $x_{0}^{2}$ in Eq.(\ref{eq5})) from the second derivative that participates to yield our relevant secular term $a^{3}\sin\phi$. Thus,
\begin{eqnarray}
&-&y_{0}x_{1}\frac{\partial^{2}f}{\partial x^{2}}(x_{0},y_{0})\Rightarrow \left[-a\sin\phi\right].\left[x_{1}\right]2f_{20}\nonumber\\
&\Rightarrow& a^{3}\sin\phi\left[\frac{1}{24}f_{xx}(7g_{xx} + 5g_{yy} - 4f_{xy})\right].\label{eqA.18}
\end{eqnarray}
\noindent Similarly, for the other three terms of type-\textbf{(ii)} we have:-\\
\begin{eqnarray}
&-&y_{0}y_{1}\frac{\partial^{2} f}{\partial x\partial y}(x_{0},y_{0})\Rightarrow\left[-a\sin\phi\right]\left[y_{1}\right]f_{11}\nonumber\\
&\Rightarrow& a^{3}\sin\phi\left[-\frac{1}{24}f_{xy}(7f_{xx} + 5f_{yy} +4g_{xy})\right]\label{eqA.19}
\end{eqnarray}

\begin{eqnarray}
&&x_{0}x_{1}\frac{\partial^{2}f}{\partial x\partial y}(x_{0},y_{0})\Rightarrow \left[a\cos\phi\right].\left[x_{1}\right]f_{11}\nonumber\\
&\Rightarrow& a^{3}\sin\phi\left[\frac{1}{12}f_{xy}(f_{xx} -f_{yy} +g_{xy}\right]\label{eqA.20}
\end{eqnarray}
\noindent and\\
\begin{eqnarray}
&&x_{0}y_{1}\frac{\partial^{2}f}{\partial y^{2}}(x_{0},y_{0})\Rightarrow\left[a\cos\phi\right][y_{1}]2f_{02}\nonumber\\
&\Rightarrow& a^{3}\sin\phi\left[\frac{1}{12}f_{yy}(-f_{xy} +g_{xx}-g_{yy})\right].\label{eqA.21}
\end{eqnarray}
\noindent Again, adding up Eq.(\ref{eqA.18}) to Eq.(\ref{eqA.21}) we get the relevant secular contributions from the type \textbf{(ii)} terms as,\\
\begin{eqnarray}
&-&y_{0}x_{1}\frac{\partial^{2}f}{\partial x^{2}}(x_{0},y_{0})-y_{0}y_{1} \frac{\partial^{2}f}{\partial x\partial y}(x_{0},y_{0})\nonumber\\
&-&x_{0}x_{1}\frac{\partial^{2}f}{\partial x\partial y}(x_{0},y_{0}) + x_{0}y_{1}\frac{\partial^{2}f}{\partial y^{2}}(x_{0},y_{0})\nonumber\\
&\Rightarrow&a^{3}\sin\phi\left[\frac{1}{24} f_{xx}(7g_{xx} + 5g_{yy}) -\frac{3}{8}f_{xy}(f_{xx}+f_{yy})\right.\nonumber\\
&+& \left.\frac{1}{12}(f_{yy}g_{xx}- f_{xy}g_{xy} - f_{yy}g_{yy})\right].\label{eqA.22}
\end{eqnarray}
\noindent The last two terms of Eq.(\ref{eq16}) consists of products of the nonlinear functions and their first derivatives. We elucidate the term $f(x_{0},y_{0})\frac{\partial f}{\partial x}(x_{0},y_{0})$ and the other one follows similarly.\\
\noindent Writing\\
\begin{equation}
f(x_{0},y_{0})\frac{\partial f}{\partial x}(x_{0},y_{0})=\sum_{i,j}\sum_{k,l} kf_{ij}f_{kl}x_{0}^{i+k-1} y_{0}^{j+l}\label{eqA.23}
\end{equation}

\noindent we again search for terms $x_{0}^{2}y_{0}$ and $y_{0}^{3}$ which give the secular $a^{3}\sin\phi$. Since ($i+j\geq2$) as well as ($k+l\geq2$), for the term $x_{0}^{2}y_{0}$ (for which $i+k=3$ and $j+l=1$), the permissible ($i~j~k~l$) combinations are ($2~0~1~1$) and ($1~1~2~0$) which add up to yield
\begin{eqnarray}
&\Rightarrow&(1.f_{20}f_{11} + 2f_{11}f_{20})x_{0}^{2}y_{0}\nonumber\\
&\Rightarrow& a^{3}\sin\phi\frac{3}{8}f_{xx}f_{xy}.\label{eqA.24}
\end{eqnarray}
\noindent For the $y_{0}^{3}$ term (where, from Eq.(\ref{eqA.23}), we have $i+k=1$ and $j+l=3$) the allowed ($i~j~k~l$) combinations are ($0~2~1~1$) and ($1~1~0~2$) which add up to give\\
\begin{eqnarray}
&\Rightarrow&(1.f_{02}f_{11} + 0.f_{11}.f_{02})y_{0}^{3}\nonumber\\
&\Rightarrow& a^{3}\sin\phi. \frac{3}{8}f_{yy}f_{xy}\label{eqA.25}
\end{eqnarray}
\noindent Adding Eq.(\ref{eqA.24}) and Eq.(\ref{eqA.25}) we get the relevant secular contribution from this term as\\
\begin{equation}
f(x_{0},y_{0})\frac{\partial f}{\partial x}(x_{0},y_{0})\Rightarrow a^{3}\sin\phi\left[\frac{3}{8}f_{xy}(f_{xx}+f_{yy})\right]\label{eqA.26}
\end{equation}

\noindent Similarly, for the term \\
\begin{eqnarray}
&&g(x_{0},y_{0})\frac{\partial f}{\partial y}(x_{0},y_{0})\nonumber\\
&=&\sum_{ij}\sum_{kl}lf_{kl}g_{ij}x_{0}^{i+k}y_{0}^{j+k-1}\label{eqA.27}
\end{eqnarray}

\noindent we get $x_{0}^{2}y_{0}$ from the ($i~j~k~l$) combinations given by ($2~0~0~2$) and ($1~1~1~1$) while the $y_{0}^{3}$ term is obtained from the combination ($0~2~0~2$). Adding them we get the $a^{3}\sin\phi$ term as
\begin{eqnarray}
&&g(x_{0},y_{0})\frac{\partial f}{\partial y}(x_{0},y_{0})\nonumber\\
&\Rightarrow& a^{3}\sin\phi \left[\frac{1}{8}f_{yy}(g_{xx}+3g_{yy})+\frac{1}{4}f_{xy}g_{xy}\right].\label{eqA.28}
\end{eqnarray}
\noindent Thus the secular $a^{3}\sin\phi$ terms at second order Eq.(\ref{eq16}) can be obtained by adding Eq.(\ref{eqA.17}), Eq.(\ref{eqA.22}), Eq.(\ref{eqA.26}) and Eq.(\ref{eqA.28}) to yield Eq.(\ref{eq23}), and hence Eq.(\ref{eqA.24}).


\end{document}